\documentclass[twocolumn,showpacs,prb,amssymb,amsmath]{revtex4}
\usepackage[dvips]{graphicx} \usepackage{dcolumn} \usepackage{bm}

\begin{document}

\title{The spectra of mixed $^3$He-$^4$He droplets} \author{S.
  Fantoni} \affiliation{SISSA and INFM DEMOCRITOS National Simulation
  Center, Via Beirut 2-4, I-34014 Trieste, Italy} %
\author{R. Guardiola} \affiliation{ Departamento de F\'{\i}sica
  At\'omica y Nuclear, Facultad de F\'{\i}sica, E-46100-Burjassot,
  Spain} \author{J. Navarro} \affiliation{ IFIC (CSIC-Universidad de
  Valencia), Apartado Postal 22085, E-46071-Valencia, Spain}
\author{A. Zuker} \affiliation{Institute de Recherches Subatomiques,
  IN2P3-CNRS/Université\'e Louis Pasteur, F-67037 Strasbourg, France }

\date{\today}

\begin{abstract}
  The diffusion Monte Carlo technique is used to calculate and analyze
  the excitation spectrum of $^3$He atoms bound to a cluster of $^4$He
  atoms, by using a previously determined optimum filling of
  single-fermion orbits with well defined orbital angular momentum
  $L$, spin $S$ and parity quantum numbers.  The study concentrates on
  the energies and shapes of the three kinds of states for which the
  fermionic part of the wave function is a single Slater determinant:
  maximum L or maximum S states within a given orbit, and fully
  polarized clusters.  The picture that emerges is that of systems
  with strong shell effects whose binding and excitation energies are
  essentially determined by averages over configuration at fixed
  number of particles and spin, {\it i.e.,} by the monopole properties
  of an effective Hamiltonian.
\end{abstract}


\maketitle

\section{introduction}\label{sec:introduction}
The study of liquid helium in confined geometries is currently an
active area of experimental and theoretical
research.~\cite{jcp01,krot02}  Helium droplets are weakly bound
quantum systems, as a consequence of their small atomic mass and the
particular form of the associated van der Waals interaction. For
$^4$He, clusters are bound for any number of atoms, while for $^3$He,
it takes about $30-32$ atoms~\cite{bar97,guar05} to form a bound
system, due to the larger zero-point motion and the Pauli Principle.
The case of mixed $^3$He-$^4$He clusters is very interesting, since
they are made of particles with different statistics and masses
interacting through the same potential. The theoretical calculations
predict the existence of instability islands for a sufficiently small
number of $^4$He atoms.\cite{guar02,bres03,bres02,guar03}

Experimentally, small helium clusters are produced by free jet
expansion of the gas. Their mass is then measured by diffraction
through a transmission grating followed by a mass spectrometer
detector.~\cite{scho94} Pure $^4$He clusters, containing up to eight
atoms, and mixed clusters, containing one $^3$He and up to six $^4$He
atoms, have been detected using a grating with 100~nm
period,~\cite{scho96} and even the very weakly bound dimer $^4$He$_2$
has been unambiguously detected.~\cite{gris00}  The experimental
setup of Ref.~\onlinecite{scho94} has been improved to detect droplet
sizes up to 25-30 amu, and very small mixed systems have been
definitely identified.~\cite{toen03}

There have been several theoretical studies of a single $^3$He atom in
a medium size $^4$He cluster, either through a density functional
approach~\cite{dalf89,barr97} or microscopic
methods.~\cite{beli94,bres00,krot01} All $^4$He$_{N_B}$$^3$He clusters
form bound states for $N_B \ge 2$. The excess in kinetic energy pushes
the $^3$He atom to the surface, resulting in a a quasi two-dimensional
wave function similar to the Andreev state describing one $^3$He
impurity in $^4$He bulk. Recently, the ordering of the single-particle
orbital states has been established.~\cite{fant04}  Larger
combinations of $^3$He and $^4$He atoms have been studied employing a
non-local finite range density functional.~\cite{barr97,pi99,nava04}
Previous variational microscopic studies of these mixed systems have
been carried out for the $^4$He$_2$$^3$He$_2$ cluster,~\cite{naka79}
and for clusters with $N_B \le 8$, $N_F \le 20$.~\cite{guar02} More
recently, the diffusion Monte Carlo (DMC) method has been
applied to droplets with $N_F \le 3$, $N_B \le 17$ in
Refs.~\onlinecite{bres02,bres03}, and with $N_F \le 20$, $N_B \le 8$
in Ref.~\onlinecite{guar03}.  The most important result from
microscopic calculations~\cite{guar02,guar03} is the prediction of
instability regions, specially when the number of $^4$He atoms is
small; nevertheless, a core with five or more $^4$He atoms is able to
bind any number $N_F$ of $^3$He atoms.

Our purpose is to analyze the ground state and the low-lying excited
states of the $^3$He atoms in a mixed helium droplet, by using the DMC
method, and relying on the single-particle orbital orderings obtained
previously~\cite{fant04} from the study of a single fermion. The
resulting spacings resemble those of the rotational spectrum of a
diatomic molecule, where the $^4$He core plays the role of one atom
and the $^3$He is the other atom. At low energies each level is
uniquely classified by its angular momentum, but vibrational-like
excitations appear at higher energies for heavy enough clusters. The
adopted filling order for large number of $^3$He atoms is therefore
$1s\ 1p\ 1d\ \dots$, with some attention being paid to a possible $2s$
intruder.  We shall concentrate on clusters with eight and twenty
bosons, varying the number of fermions. This moderately large number
of bosons ensures the existence of several bound excited levels.  All
calculations have been made using the HFD-(B) potential of Aziz and
co-workers~\cite{HFD-B} for the He-He interaction.

The plan of the paper is as follows. In Section II we give some general
details concerning the DMC calculation.  Section III collects
single-fermion results needed later.  Section IV deals with the case
of two fermions: this system is simple enough to be analyzed in depth
for a large set of quantum numbers, shedding light in particular on
the single-particle ordering.
Section V is devoted to stable mixed clusters, for the specific cases
of maximum spin and maximum orbital angular momentum. The case of
fully polarized clusters is also considered.
In Section VI we exhibit the calculated one- and two-body
distribution functions, both for normal and fully polarized mixed
clusters. Finally, in Section VII we summarize our findings and draw some
general conclusions.

\section{The diffusion Monte Carlo method}\label{sec:diff-monte-carlo}
The DMC~\cite{Reyn82,Mosk82} method is based on an importance sampling
wave function, with the double role of controlling the variance of the
ground state energy and to incorporate both the required statistics
and the desired spin and angular momentum quantum numbers. In this
work we use the same form employed previously.~\cite{guar03} The trial
wave function is written as the product of four factors,
\begin{equation}\label{is}
  \Psi_T = \Phi_{BB} \Phi_{BF} \Phi_{FF} D_F
\end{equation}
corresponding to a Jastrow form for the boson-boson $(BB)$,
boson-fermion $(BF)$, and fermion-fermion $(FF)$ parts, and the Slater
determinantal part $D_F$ for the fermions.  Each Jastrow term is a
translationally invariant and symmetric wave function, with the
structure
\begin{equation}\label{jastrow}
  \Phi_{MN} =
  \prod _{i,j} e^{f_{MN}(r_{ij})}
\end{equation}
where indices $M,N$ represent bosons $(B)$ or fermions $(F)$, and
indices $i,j$ run over the corresponding atoms. The product runs over
all different pairs $i<j$ if $M=N$, and no restrictions apply when
$M\neq N$. We have used a simple but nevertheless physically complete
representation
\begin{equation}\label{ff}
  f_{MN}(r) = - \frac{1}{2} \left( \frac{b_{MN}}{r}\right)^{\nu_{MN}}  -
  r\, p_{MN},
\end{equation}
depending on three parameters, $b$, $p$ and $\nu$, in general
different for each of the three $BB$, $BF$ and $FF$ pairs. The short range
coefficients $b$ and $\nu$ have been fixed, independently of the
number of bosons or fermions, to the values $\nu = 5.2$ for any subset
$MN$, $b_{BB}=2.95$~\AA{}, $b_{BF}=2.90$~\AA{}, and
$b_{FF}=2.85$~\AA{}. The long range part $r\, p_{MN}$ entering
Eq.~(\ref{ff}), has the role of confining the system and fixing
roughly its size. The three parameters $p_{MN}$ are adjusted for each
droplet by minimizing its ground state energy.

The last factor $D_F$ in the trial wave function is the determinantal
part, which will be discussed later on for each specific case. The
description of the trial wave function is completed by the inclusion
of the Feynman-Cohen backflow~\cite{feyn56} in the fermionic
exponential tail as well as in the radial dependence of the Slater
factor. Following Ref.~\onlinecite{schm81}, we have replaced each
fermion coordinate ${\bf r}_i$ by a transformed coordinate
\begin{equation}
  \label{backflow}
  \hat{\mathbf r}_i = {\mathbf r}_i + \sum_{i \neq j} \eta(r_{ij})
  \left( {\mathbf r}_i - {\mathbf r}_j \right)
\end{equation}
For the backflow function $\eta(r)$ we choose the medium-range form
used in Ref.~\onlinecite{pand86}:
\begin{equation}
  \eta(r) = \frac{\lambda}{r^3}
\end{equation}
keping the same value for the parameter $\lambda=5$~\AA$^3$.

Using the model described above for the importance sampling guiding
function, a real-time DMC evolution has been carried out employing a
${\cal O}(\tau ^3)$ approximation to the Green function, with time slice
$\tau=0.00025\ $K$^{-1}$, for a total of 200 blocks of 200 steps each.
A block average was used in order to diminish the unavoidable
correlations of the DMC method, with the aim of obtaining a reasonable
estimate of the variance. An initial population of 1000 walkers lead,
on the average, to a total number of forty millions samples.

In spite of this block averaging, the resulting estimate of the
variance turned out to be quite optimistic, {\em i.e.,} much smaller
than reasonably expected. As a consequence, we opted for a
very costly but safe procedure, by carrying out ten independent
calculations with the same time slice and the same number of samples,
but with randomly selected initial set of walkers. The resulting ten
results lead to the {\em true} variance (or so we hope). The numbers
quoted in what follows correspond to this prescription.

Due to the presence of the Slater determinants, the importance
sampling function is not definite positive. In our calculations, the
random process has been constrained using the so-called fixed node
approximation: any walker attempting to cross a nodal surface is
neglected. As it has been shown,~\cite{mosk82,reyn82} the use of this
approximation leads to an upper bound to the lowest energies.

\section{The one fermion system}\label{sec:one-fermion-system}
The system made of a single $^3$He atom plus a drop of $^4$He atoms
has been recently investigated.~\cite{fant04} For the sake of
completeness we summarize in this section the most relevant results
for the analysis of systems with more fermions.

As there is only one fermion, the Slater determinant in Eq.~(\ref{is})
becomes a single-particle wave function $\phi_{n\ell m}({\bf r})$ of
radial quantum number $n$ and orbital angular momentum $\ell$ with
projection $m$.  Translational invariance is ensured by referring the
fermion coordinate to the bosonic center of mass.

Different values of $\ell$ select a specific angular momentum subspace
in which the DMC procedure will drive the wave function so as to
minimize the energy.  Thus, the radial part of the fermionic wave
function can be taken to have the form
\begin{equation}
  \label{sp-wf}
  \phi_{1\ell m} (r) = r^\ell Y_{\ell m}(\hat r),
\end{equation}
which ensures that there are no radial nodes ($n=1$). These
single-particle wave functions do not contain a radial confining
term, because it is already included in the Jastrow part of the
importance sampling wave function.  Notice that within the DMC
procedure it is not possible to obtain excited levels with $n>1$,
because this will require to impose a strict orthogonality on the
physical $n=0$ ground state. Nevertheless, one may use an indirect
procedure based in the moment method which will provide an upper bound
to the energy of the radial excited levels.

Table~\ref{previous-calc} displays the energy of these one-fermion
states, as a function of the single-fermion quantum numbers.  The
quoted energies are slightly different from those of Ref.~\onlinecite{fant04}
because of the improved statistics of the present calculation.  The
ground state energies of the pure bosonic droplets are also given, so
as to define the {\em separation energies}
\begin{equation}
  \label{sep-energy}
  \epsilon_{n\ell} = E_{n\ell}(N_B, N_F=1) - E (N_B, N_F=0),
\end{equation}
also quoted in the Table. Positive values refer to unbound levels.
The energy obtained for the $n=2,\ell=0$ excited level is above the
dissociation limit for $N_B=8$ or very close to it for $N_B=20$.

\begin{table}[ht!]
  \caption{Energies  of droplets with eight and twenty bosons and one fermion,
    for several values of the angular momentum. The row labelled
    $N_F=0$ is the system without fermions. The columns labelled
    $\epsilon$ are the separation energies.
    Values  in italics correspond to unbound levels.}
  \label{previous-calc}
  \begin{tabular}{|l|ll|ll|}
    \hline
    Config & \multicolumn{2}{|c|}{$N_B=8$} & \multicolumn{2}{|c|}{$N_B=20$}\\
    \hline
    & E(K)       & $\epsilon$(K) & E(K)  & $\epsilon$ (K) \\
    \hline
    $N_F=0$ & -5.14(1) &       & -33.76(2) & \\
    $1s$    & -6.08(1) & -0.94 & -35.55(1) & -1.79\\
    $1p$    & -5.60(1) & -0.46 & -35.15(2) & -1.39\\
    $1d$    & {\em -4.98(1)}   & 0.16      & -34.55(2) & -0.89 \\
    $2s$    & {\em -5.10    }  & 0.06      & -33.80 & -0.04 \\
    \hline
  \end{tabular}
\end{table}

The single-particle wave functions of Eq. (\ref{sp-wf}) will be used
later to construct model wave functions for systems with two or more
fermions.  Unfortunately the moment method does not provide a wave
function for the radial excitation, and for the calculations of the
following sections we will use the simple form
\begin{equation}
  \label{sp-wf-2s}
  \phi_{n=2,\ell=0 } (r) = r^2,
\end{equation}
without any radial node. The lack of nodes may be a deficiency, but
the truly important point is that the simple form chosen is linearly
independent of the $1s$ state.

\begin{table}[hb!]
  \caption{Space part of importance sampling singlet(upper part) and
    triplet (lower part)
    wave functions for the two-fermion drop.}
  \label{singlet-wf}
  \begin{tabular}{llll}
    \hline
    $L^P$& Config & $\Psi$ (SINGLET)\\
    &      &        &       \\
    $ 0^+$&$1s^2$ & 1 \\
    $ 0^+$&$1p^2$ & ${\bf r}_1 \cdot {\bf r}_2$\\
    $ 0^+$&$1d^2$ & $ 3 ({\bf r}_1\cdot{\bf r}_2)^2 - r_1^2 r_2^2$ \\
    $ 1^-$&$1s1p$ & $z_1 + z_2$\\
    $ 1^-$&$1p1d$ & $ z_1 r_2^2 - 3 z_2 ({\bf r}_{1}\cdot {\bf r}_2)
    z_2 r_1^2 - 3 z_1 ({\bf r}_{1}\cdot {\bf r}_2)$ \\
    $ 2^+$&$1s1d$ & $x_1^2 + x_2^2 -y_1^2 - y_2^2$\\
    $ 2^+$&$1p^2$ & $ x_1 x_2 - y_1 y_2$ \\
    $ 2^+$&$1d^2$ & $(x_1^2-y_1^2) r_2^2 +(x_2^2-y_2^2)r_1^2 - $\\
    &       & $ 3(x_1x_2-y_1y_2) ({\bf r}_1\cdot {\bf r}_2)$ \\
    $ 2^-$&$1p1d$ & $ x_1 x_2 z_2 - x_2^2  z_1 - y_2 (y_1 z_2 - y_2 z_1)+$   \\
    &      & $ x_2 x_1 z_1 - x_1^2  z_2 - y_1 (y_2 z_1 - y_1 z_2)$\\
    $ 3^-$&$1p1d$ & $ x_1 (y_2^2  - x_2^2 ) + 2 x_2 y_1 y_2 +$  \\
    &      & $x_2 (y_1^2  - x_1^2 ) + 2 x_1 y_1 y_2$ & \\
    $4 ^+$&$1d^2 $& $ (x_1^2- y_1^2)(x_2^2-y_2^2) - 4 x_1 x_2 y_1 y_2$ \\
    $0 ^+$&$1s2s$ & $r_1^2+r_2^2$ \\
    $1 ^-$&$1p2s$ & $z_1 r_2^2+z_2 r_1^2$ \\
    \hline
    \hline
    $ L^P$ &Config & $\Psi$ (TRIPLET) \\
    &       &       \\
    $1 ^-$&$1s1p$ &  $z_1 - z_2$\\
    $1 ^-$&$1p1d$ &  $ z_1 r_2^2 - 3 z_2 ({\bf r}_{1}\cdot {\bf r}_2) -
    z_2 r_1^2 + 3 z_1 ({\bf r}_{1}\cdot {\bf r}_2)$\\
    $1 ^+$&$1p^2$ &  $x_1 z_2 - x_2 z_1$\\
    $1 ^+$&$1d^2$ &  $ (x_1 z_2-x_2 z_1)({\bf r}_1 \cdot {\bf r}_2)$\\
    $2 ^+$&$1s1d$ &  $x_1^2 - x_2^2 -y_1^2 + y_2^2$\\
    $2 ^-$&$1p1d$ &  $ x_1 x_2 z_2 - x_2^2  z_1 - y_2 (y_1 z_2 - y_2 z_1)-$ \\
    &       & $ x_2 x_1 z_1 + x_1^2  z_2 + y_1 (y_2 z_1 - y_1 z_2)$  \\
    $3 ^+$&$1d^2$ &  $ x_1^2(y_2^2-z_2^2) - x_2^2(y_1^2-z_1^2) + y_1^2 z_2^2-y_2^2 z_1^2$
    \\
    $3 ^-$&$1p1d$ &  $ x_1 (y_2^2  - x_2^2 ) + 2 x_2 y_1 y_2 -$\\
    &       & $x_2 (y_1^2  - x_1^2 ) - 2 x_1 y_1 y_2 $ \\
    $0 ^+$ & $1s2s$ & $r_1^2 - r_2^2$ \\
    $1 ^-$ & $1p2s$ & $z_1 r_2^2 - z_2 r_1^2$ \\
    \hline
  \end{tabular}
\end{table}

\section{The two fermion system}\label{sec:two-fermion-system}

Adding two $^3$He atoms to a core of $^4$He atoms results in a system
deeply resembling the helium atom. To a large extent, the bosonic
sub-cluster plays the role of the atomic nucleus, with the two $^3$He
atoms corresponding to the electrons. There are two families of levels,
singlet ($S=0$) and triplet ($S=1$).  Each of the states is
characterized by the configuration, the orbital angular momentum $L$,
the spin $S$ and the parity $P$.

Specific two-fermion states are constructed from the single-particle
wave functions~(\ref{sp-wf}) by coupling the angular momentum part to
the required quantum numbers and by symmetrizing the singlet or
antisymmetrizing the triplet radial wave functions. In general one does not
obtain real wave functions, but the remedy is simple. For $M=0$ the
result is real. Otherwise construct two cases for $M$ and $-M$ and
either add or subtract them. The wave function thus constructed has no
good $L_z$, but still good total orbital angular momentum, and should
have no effect on the computed energy.

For the sake of completeness we list in Table~\ref{singlet-wf} the
specific forms thus obtained, and used as importance-sampling wave
functions to drive the DMC stochastic procedure.

\begin{table}[htb]
  \caption{Binding energies, in K, for several
    states with two fermions and  $N_B=8$ (upper table) and $N_B=20$ (lower
    table), classified by the
    configuration and angular momentum quantum numbers. Values in bold
    face correspond
    to the physically interesting
    states. The dissociation limit is 6.08 K ($N_B=8$) and
    35.55 K ($N_B=20$).
    The statistical errors of the energies are between 0.01  and 0.02 K.
    The row labelled {\em Eff.} is the prediction of the
    non-interacting fermion model.
  }
  \label{two-fermion-8}
  \begin{tabular}{|c|c|c|c|c|c|c|c|c|}
    \hline
    \multicolumn{9}{|c|}{$N_B=8$}\\
    \hline
    $L^PS$ & $1s^2$   & $1s1p$      & $1p^2$       & $1s1d$ & $1p1d$ & $1d^2$ & $1s2s$ & $1p2s$ \\
    \hline
    $0^ +$0 & {\bf 7.09}&             & 6.12         &        &        & 4.90   & 7.07   &        \\
    $1^ -$0 &       & {\bf 6.61}      &              &        & 5.49   &        &        & 6.49   \\
    $2^ +$0 &       &                 & {\bf 6.15}   & 5.99   &        & 4.88   &        &        \\
    $2^ -$0 &       &                 &              &        & 5.50   &        &        &        \\
    $3^ -$0 &       &                 &              &        & 5.48   &        &        &        \\
    $4^ +$0 &       &                 &              &        &        & 4.85   &        &        \\
    $0^ +$1 &       &                 &              &        &        &        & 5.89   &        \\
    $1^ +$1 &       &                 & {\bf 6.19}   &        &        & 5.06   &        &   \\
    $1^ -$1 &       & {\bf 6.65}      &              &        & 5.65   &        &        & 6.46   \\
    $2^ +$1 &       &                 &              & 6.01   &        &        &        &        \\
    $2^ -$1 &       &                 &              &        & 5.52   &        &        &        \\
    $3^ +$1 &       &                 &              &        &        & 4.89   &        &        \\
    $3^ -$1 &       &                 &              &        & 5.45   &        &        &        \\
    \hline
    {\em Eff.} &
    7.02 & 6.54 & 6.06 & 5.92 & 5.44 & 4.82 & 6.02 & \\
    \hline
    \hline
    \multicolumn{9}{|c|}{$N_B=20$}\\
    \hline
    $L^PS$  & $1s^2$ & $1s1p$ & $1p^2$ & $1s1d$ & $1p1d$ & $1d^2$ & $1s2s$ & $1p2s$ \\
    \hline
    $0^ +$0 &{\bf 37.33}  &              & 36.70        &              &              & 35.45  & 37.33  &        \\
    $1^ -$0 &             & {\bf 37.06}  &              &              & 36.05        &        &        & 36.98  \\
    $2^ +$0 &             &              & {\bf 36.74}  & 36.47        &              & 35.54  &        &        \\
    $2^ -$0 &             &              &              &              & {\bf 36.19}  &        &        &        \\
    $3^ -$0 &             &              &              &              & {\bf 36.17}  &        &        &        \\
    $4^ +$0 &             &              &              &              &              & 35.55  &        &        \\
    $0^ +$1 &             &              &              &              &              &        & {\bf 35.68}  &  \\
    $1^ +$1 &             &              & {\bf 36.74}  &              &              & 35.79  &        &       \\
    $1^ -$1 &             & {\bf 37.05}  &              &              & 36.27        &        &        & 36.99  \\
    $2^ +$1 &             &              &              & {\bf 36.49}  &              &        &        &        \\
    $2^ -$1 &             &              &              &              & {\bf 36.20}  &        &        &        \\
    $3^ +$1 &             &              &              &              &              & 35.66  &        &        \\
    $3^ -$1 &             &              &              &              & {\bf 36.15}  &        &        &        \\
    \hline
    {\em Eff}.
    & 37.34        & 36.94         & 36.54         & 36.44         &  36.04        & 35.44 & 35.59 & \\
    \hline
  \end{tabular}
\end{table}

The energies obtained from these configurations are displayed in Table
\ref{two-fermion-8}, for $N_B=8$ and $N_B=20$.  Some among these results
lie above the dissociation limit ({\em i.e.,} the energy of the lowest
bound state with the same number of bosons but with a single fermion)
and do not correspond strictly to truly bound systems.

It has to be understood that in the results quoted in these two
tables, only a single number per row does have a physical sense: the DMC
algorithm improves systematically the importance sampling wave
function, but because of the use of the fixed-node approximation, that
improvement only provides a variational upper bound to the energy for
each of the subspaces with well defined $L$, $S$ and $P$ quantum
numbers. To give an example: the state $^1S$ may have projections on
the $1s^2$, $1p^2$, $1d^2$ and other shell-model states, but the
mixing will probably not be constructed along the DMC stochastic
procedure. Thence, the DMC physically relevant results are those with
larger value of the binding energy for each row.  The other
configurations with smaller energy are possible interacting
configurations, and presumably an optimized linear combination within
each row could provide a better binding energy.

\begin{figure}
  \caption{The two-fermion spectrum for $N_B=8$ and $N_B=20$,
    classified accordingly to the configuration. Energies are in
    Kelvin.  The horizontal line is the dissociation limit. Note that
    apart from the energy shift, the scales of the two plots are the
    same.}
  \label{twoferm8}
   \includegraphics[width=8cm]{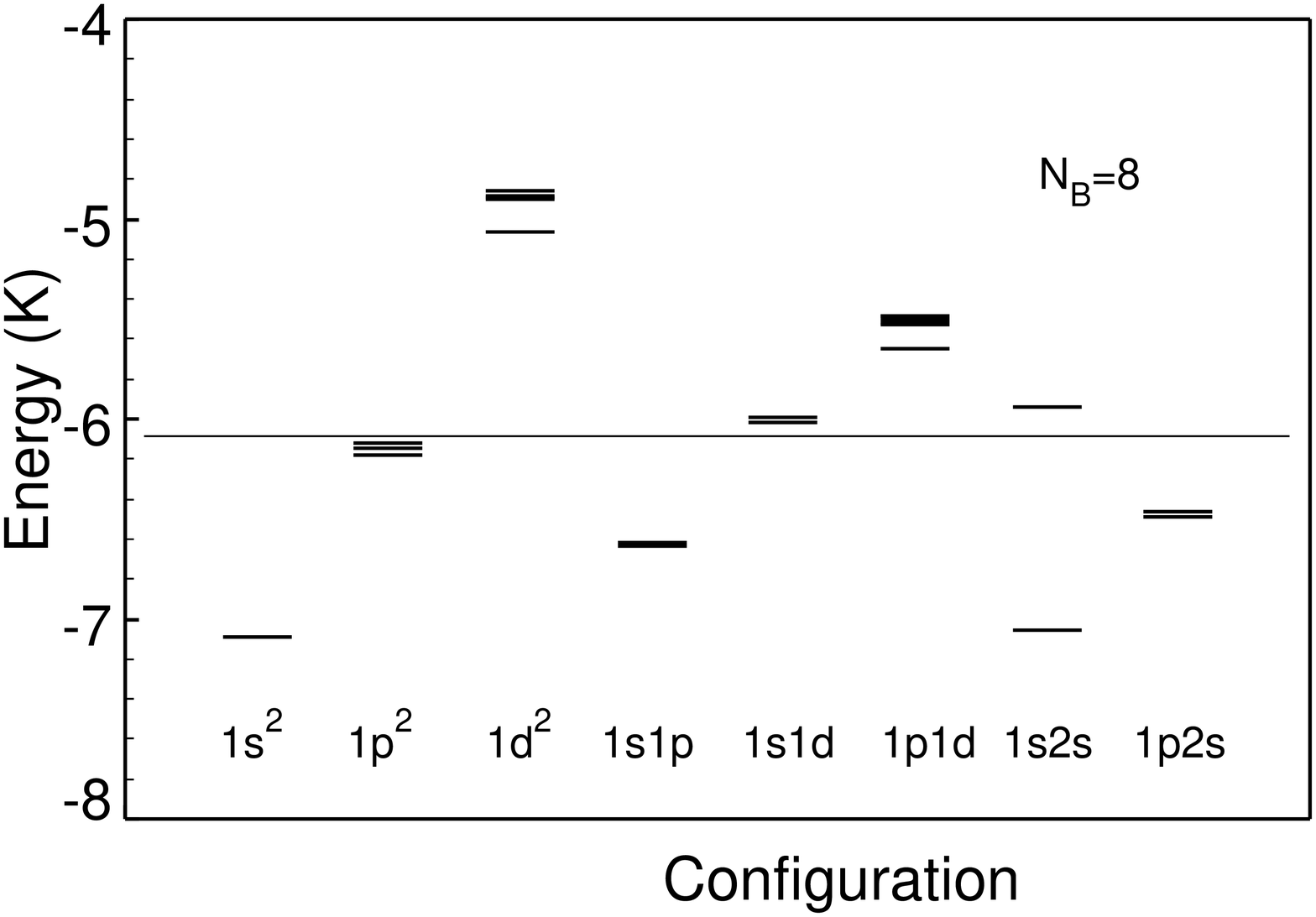}
   \\
   \includegraphics[width=8cm]{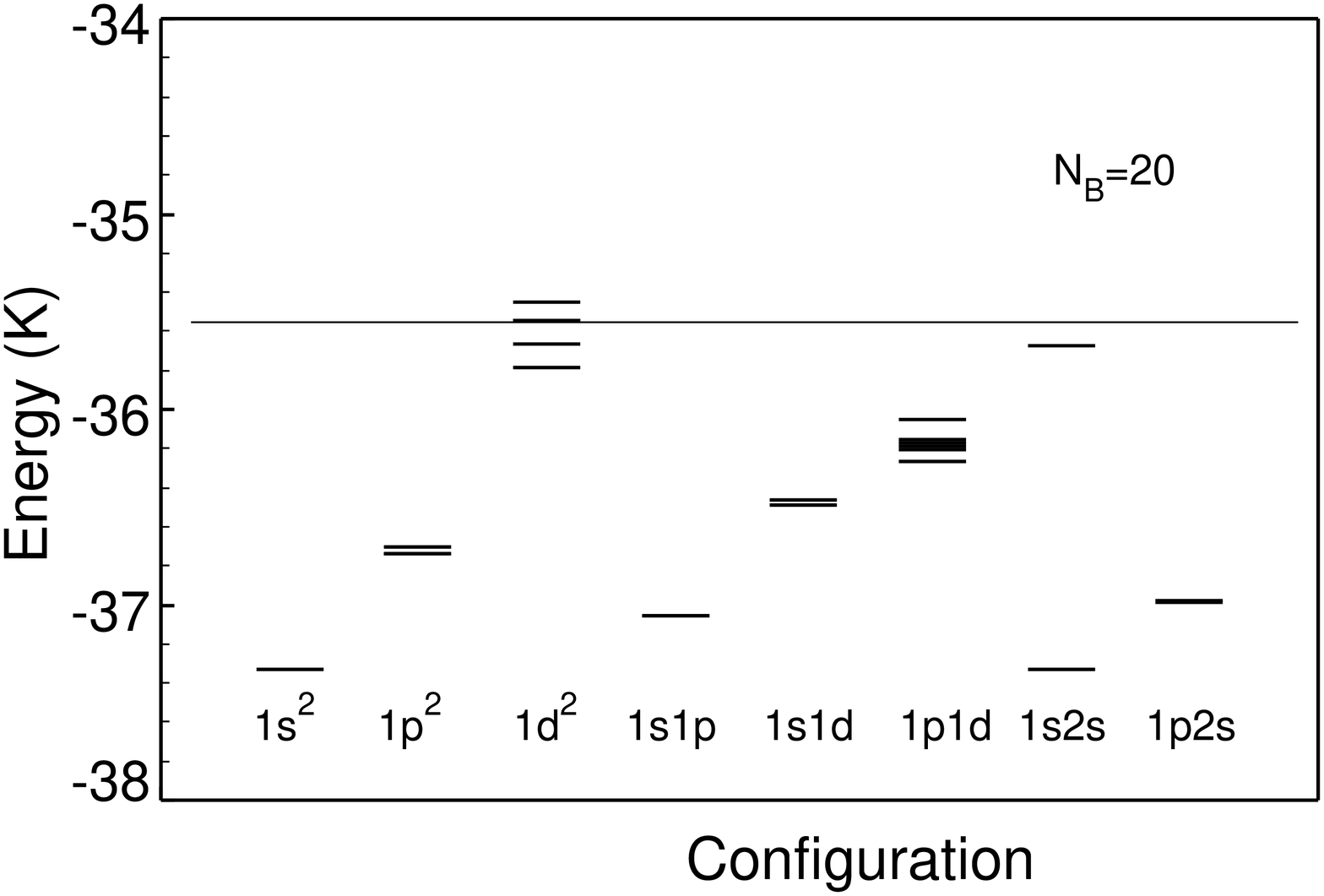}
\end{figure}

\begin{figure}
  \caption{The root mean square radius (in \AA) of bosons and fermions
    for two-fermion systems, referred to the full center-of-mass of
    the drop, as a function of the configuration. The two plots
    correspond to $N_B=8$ (upper figure) and $N_B=20$ (lower figure).
    The lines displaying an almost horizontal lines represent the
    boson radii, and the varying lines represent the fermion radii.}
  \label{radius8}
   \includegraphics[width=8cm]{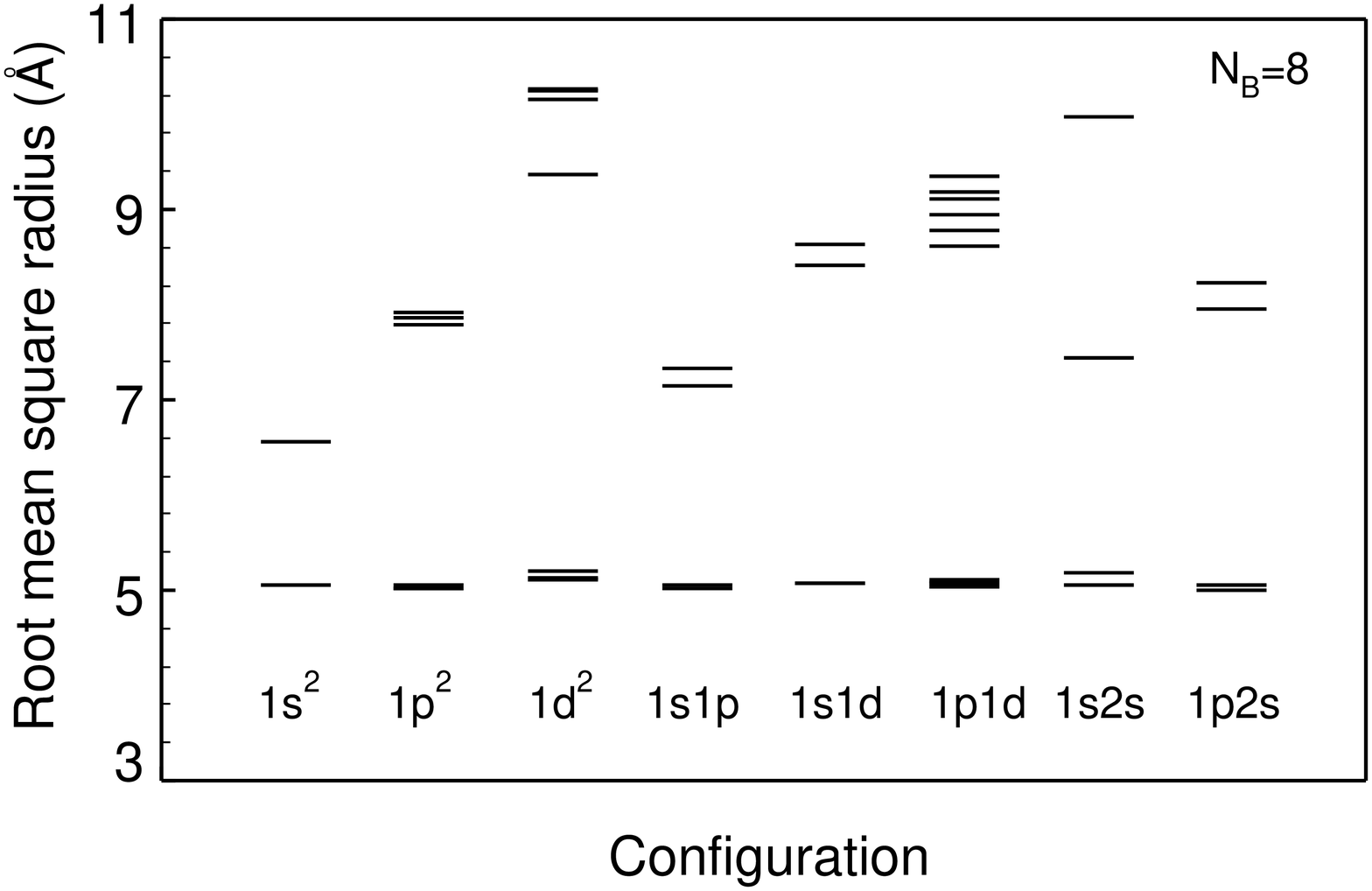}
   \\
   \includegraphics[width=8cm]{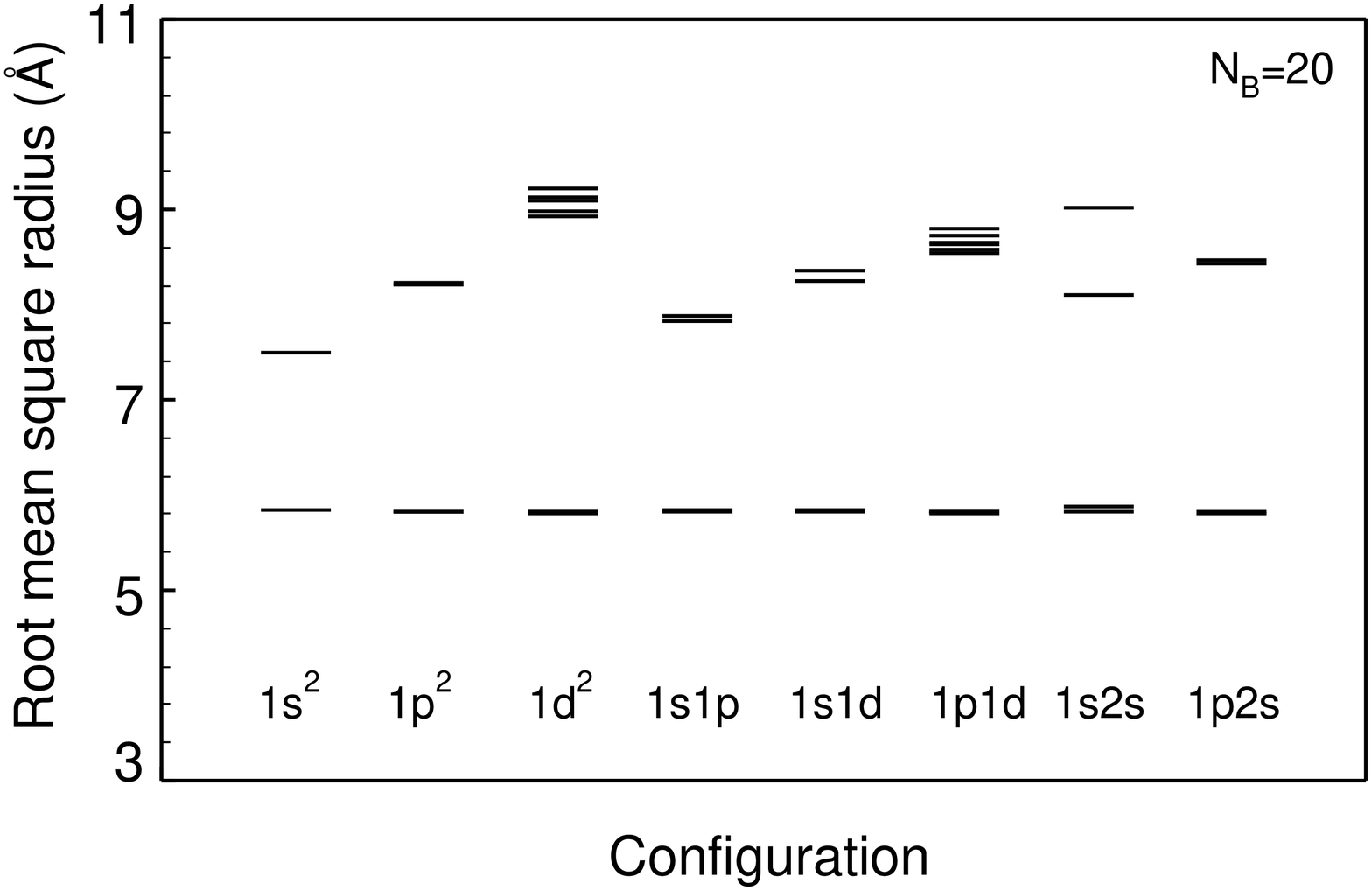}
\end{figure}

Note that among the states involving the $2s$ shell none of them has maximum
binding except $1s2s:^3\!\!S$ because the configuration is unique.
For $N_B=8$ this level is not bound, but it is the last bound one for
$N_B=20$. For the other cases the DMC optimized energies prefer $1s$
over $2s$ orbits, but $1s2s:^3\!\!S$ survives because $1s^2:^3\!\!S$
violates the exclusion principle.

Concentrating on the normal shells ($1s,\, 1p,\, 1d$) it appears that
the binding energies depend basically on the configuration, and are
almost independent of the coupling within the configuration.
Figs.~\ref{twoferm8} and~\ref{radius8} give an idea of this near
independence, both for energies and radii.  Within a central-field
shell-model description, this fact indicates that the {\em residual
  interaction} between the two $^3$He atoms is very small, with
energies close to the ones provided by the simple non-interacting
picture $$ E(n_1 \ell_1, n_2 \ell_2) = E_0 + \epsilon_{ n_1 \ell_1} +
\epsilon_{ n_2 \ell_2}, $$ where $E_0$ is the energy of the boson core
and $\epsilon_{n\ell}$ the separation energies defined in
Eq.~(\ref{sep-energy}) and given in Table \ref{previous-calc}.  The
corresponding values appear in the last row of
Table~\ref{two-fermion-8}, with the exception of the last column, the
$1p2s$ configuration which actually has evolved close to the $1s1p$
configuration.

The rough picture of the two-fermion drops as basically
non-interacting fermions bound to a rigid $^4$He cluster will be
refined in Section~\ref{sec:effect-monop-inter} by introducing an
effective monopole interaction.

\section{Clusters with more than two fermions: Binding energies}
\label{sec:clusters-with-more}

In this Section we consider selected states with a fixed number of bosons and an
increasing number of fermions, up to 18 for normal clusters, {\em i.e.},
with fermions with spin up and down, and up to 9 for fully polarized
clusters, with only spin up fermions.

\subsection{The Slater determinant}\label{sec:slater-determinant}
The fermionic factor $D_F$ must have good angular momentum and spin
quantum numbers, and it must be properly antisymmetrized and
translationally invariant. We take it to be a product of two Slater
determinants, one for each spin orientation. The obvious way to build
them up is with single-particle functions $\phi_{n \ell m}$ generated by
a central field which dictates a natural filling order.  In general,
good total orbital angular momentum $L$ and total spin $S$ demand a
linear combination of determinants. If we insist on a single product,
the construction is quite cumbersome for shells with high $\ell$ (see
{\em e.g.} Ref.~\onlinecite{cond51}), though simple when dealing with $s$
and $p$ shells.

The main problem in establishing a reasonable shell-model description
of a system containing $^3$He atoms, both in pure as well as in mixed
drops, is the lack of phenomenological information about the central
field. There are two familiar schemes common to other fermionic
systems: the shell ordering $1s,2s,2p,3s,3p \dots$, characteristic of
atoms, and the harmonic oscillator sequence $1s, 1p, (2s,1d), (2p,1f)
\dots$, used in light and medium atomic nuclei. The principal quantum
number follows different rules in both schemes. For the latter the
parenthesis indicate degenerate orbitals.

In the previous Sections we have explained why the
$1s,1p,1d,2s,\dots$ ordering was adopted for the calculations.
The information comes from
the analysis of a single $^3$He atom bound to a medium-size bosonic
drop, which has been studied by density functional
methods~\cite{dalf89, barr97} and by microscopic methods based either
in variational wave functions~\cite{beli94,krot01} or in DMC
techniques.~\cite{fant04}  All these studies suggest that the fermion
may be viewed as a particle bound to a potential well, centered in the
bosonic drop, but which is appreciably different from zero (and
attractive) only in a rather wide region near the surface of the drop,
but goes to zero near the center of the drop as well as at long
distances. This single-particle potential gives a special level
ordering based on the orbital angular momentum, $1s,1p,1d,1f \dots$,
with almost degenerate single-particle energies. Moreover, the fermion
has a very small probability penetrating the boson drop. The same
scheme results from the study of one $^3$He atom in liquid $^4$He,
giving rise to the so-called Andreev states, as well as from the study
of many $^3$He atoms attached to a large core of $^4$He
atoms.~\cite{nava04} Note, however, that {\em intruder} levels, like
$2s$, may appear for a sufficiently large number of fermions and a
sufficiently small number of bosons.

Among the possible wave functions related to a given configuration
(the so-called {\em terms} in Atomic Physics) we have chosen the two
simplest cases: a) maximum total spin on which maximum total orbital
angular momentum is built; b) maximum orbital angular momentum on
which maximum spin is built. The resulting wavefunctions are products
of two determinants, one for each spin orientation.  Other choices
demand linear combination of Slater determinants.

As the exponential tail in the trial wave function has the role of
roughly confining the system, we can construct the Slater determinants
considering only the angular momentum part of the single-particle
functions, as well as the spin part. As mentioned above, we have used
the harmonic polynomials $\phi_{\ell m}({\mathbf r}) =r^{\ell} Y_{\ell
  m}(\Omega)$ as single-particle functions. The determinants so
constructed are translationally invariant wave functions, in the sense
that they only depend on the  $3N_F-3$ relative coordinates
${\mathbf r}_i - {\mathbf  r}_j$. This fact is of crucial importance,
particularly when describing systems with a small number of constituents.

The way of constructing the required determinants is very simple. Take
for instance the maximum spin case. Once the innermost shells have
been filled, the remaining $^3$He atoms occupy the $\phi_{\ell \ell}$,
$\phi_{\ell\, \ell-1}$ \dots spin up states, until the angular
momentum states are exhausted; then the same procedure is followed to
fill out the spin down states. The spin $S$ of the resulting
determinant has the maximum value allowed for the occupancy of the
shell, and its orbital angular momentum is $L=|L_z|$. In general the
value of the determinant is a complex number, not very adequate for
the DMC algorithm. The solution is as in the two-fermion case: to use
the sum or difference (whichever is non vanishing) of the determinants
with $L_z=L$ and $L_z=-L$. The importance sampling wave function has
well defined $S$, $S_Z$ and $L$, but not $L_z$.  Nevertheless, due to
the rotational symmetry of the Hamiltonian, this has no influence on
the energy values. However the need of computating determinants with
complex matrix elements still remains, with the consequent slowing down of
the numerical calculations.

By using this procedure we have calculated states with up to $N_F=18$,
corresponding to the complete filling of the three lowest shells.  We
have also considered states in which all spins are up, {\em i.e.}, fully
polarized fermions, with the maximum number $N_F=9$. The procedure to
construct the determinants in this case is an obvious adaptation of
the one described above.

In previous work~\cite{guar02,guar03} we used a conventional {\em
  Cartesian} ordering, in particular $\{ x^2, y^2, z^2, xy,xz,yz\}$,
whereby the single-particle orbitals are a mixture of $2s$ and $1d$
wave functions. As a consequence the differences with the present
calculations---associated to changes in the nodal structures---become
significant when the $d-$shell starts to be filled.

\subsection{Binding energies of normal clusters}
\label{sec:bind-energ-norm}

Table~\ref{energiesbis} presents the values of the binding energies
corresponding to two situations, $N_B=8$ and $N_B=20$, for values of
$N_F$ from 0 to 18. All these values have been computed with the level
ordering discussed above, and for two possible couplings: $S_{max}$,
where in each shell particles are aligned to maximum spin $S$, and
then to to maximum orbital angular momentum $L$; $L_{max}$, where
particles are first aligned to maximum $L$, and then maximum $S$. In
Table~\ref{energiesbis}, whenever there are two entries for a given
cluster, the upper row corresponds to $S_{max}$, and the lower one to
$L_{max}$.  Maximum spin is quite uniformly favoured but the splitting
of the two computed levels is always smaller than 0.3-0.4~K.

\begin{table}[h]
  \caption{
    Binding energies (in K) of mixed clusters with 8 and 20 bosons and
    up to 18 fermions. Whenever there are two entries for a given
    cluster, the upper row corresponds to the $S_{max}$ coupling:
    particles aligned to maximum $S$ and then to maximum $L$. The
    lower row is for the  $L_{max}$ case: maximum $L$ first and then
    maximum $S$. In the last  column the results obtained in
    Ref.~\protect{\onlinecite{guar03}} for $N_B=8$  are also displayed.}
  \label{energiesbis}
  \begin{tabular}{r|l|ll|rr|r}
    \hline
    $N_F$& Conf     & L & S & $N_B=8$    & $N_B=20$ &
    $N_B=8$~[\protect{\onlinecite{guar03}}]\\
    \hline
    0  &          & 0 & 0 & 5.14(0)  & 33.76(1) & 5.13(2) \\
    \hline
    1  & $1s^1$   & 0 &1/2& 6.08(0)  & 35.55(1) & 6.07(2) \\
    2  & $1s^2$   & 0 & 0 & 7.09(0)  & 37.32(1) & 7.05(2) \\
    \hline
    3  & $1p^1$   & 1 &1/2& 7.72(0)  & 38.88(1) & 7.69(2) \\
    4  & $1p^2$   & 1 & 1 & 8.44(1)  & 40.47(2) & 8.42(2) \\
    &             & 2 & 0 & 8.40(1)  & 40.44(2) & \\
    5  & $1p^3$   & 0 &3/2& 9.25(1)  & 42.14(1) & 9.23(2) \\
    &             & 2 &1/2& 9.20(1)  & 42.08(2) & \\
    6  & $1p^4$   & 1 & 1 & 10.09(1) & 43.72(3) & 10.03(3) \\
    &             & 2 & 0 & 10.04(1) & 43.71(2) & \\
    7  & $1p^5$   & 1 &1/2& 11.00(1) & 45.40(2) & 11.03(3) \\
    8  & $1p^6$   & 0 & 0 & 12.00(1) & 47.07(2) & 12.03(3) \\
    \hline
    9  & $1d^1$   & 2 &1/2& 12.49(1) & 48.37(2) & 12.33(3) \\
    10 & $1d^2$   & 3 & 1 & 13.02(1) & 49.62(2) & 12.74(3) \\
    &             & 4 & 0 & 12.97(1) & 49.64(4) & \\
    11 & $1d^3$   & 3 &3/2& 13.65(1) & 51.03(4) & 13.20(4) \\
    &             & 5 &1/2& 13.56(1) & 51.01(3) & \\
    12 & $1d^4$   & 2 & 2 & 14.42(1) & 52.46(5) & 13.71(5) \\
    &             & 6 & 0 & 14.19(1) & 52.23(4) & \\
    13 & $1d^5$   & 0 &5/2& 15.26(2) & 53.99(2) & 14.20(4) \\
    &             & 6 &1/2& 14.96(2) & 53.71(3) & \\
    14 & $1d^6$   & 2 & 2 & 16.01(1) & 55.37(4) & 14.88(4) \\
    &             & 6 & 0 & 15.74(2) & 55.19(4) & \\
    15 & $1d^7$   & 3 &3/2& 16.77(2) & 56.83(3) & 15.73(5) \\
    &             & 5 &1/2& 16.64(1) & 56.70(5) & \\
    16 & $1d^8$   & 3 & 1 & 17.63(2) & 58.34(4) & 16.55(4) \\
    &             & 4 & 0 & 17.62(2) & 58.29(4) & \\
    17 & $1d^9$   & 2 &1/2& 18.64(3) & 59.94(3) & 17.44(13) \\
    18 &$1d^{10}$ & 0 & 0 & 19.74(3) & 61.56(5) & 18.49(5) \\
    \hline
  \end{tabular}
\end{table}

The present results for $N_F\leq 8$ should coincide, and indeed they
do, with previous calculations based in the cartesian ordering of the
single-particle states.~\cite{guar03} Beyond $N_F=8$ the calculations
of Ref.~\onlinecite{guar03} used an uncontrolled mixture of $1d$ and
$2s$ states, as explained above, thus corresponding to different
importance sampling functions. The last column of
Table~\ref{energiesbis} displays the results obtained in
Ref.~\onlinecite{guar03} for $N_B=8$ clusters. One should keep in mind
that both results are based on the DMC method within the fixed node
approximation, so that in both cases the obtained energies are
actually upper bounds to the real ones. The present binding energies
for $N_F>8$ are slightly higher that the previous results, the gain
being of 0.16~K for $N_F=9$ and monotonically increasing up to 1.40~K
for $N_F=18$, in the case of $N_B=8$. This apparently modest increase
(up to 6\%) may be relevant for the boundaries of the stability chart
of mixed drops. In any case, it gives support to the level ordering
used in the present work, as the associated importance sampling
function provides a better variational bound than the previous ones.

\begin{figure}[!htbb]
  \caption{Binding energies (in K) for $N_B=8$ and $N_B=20$, as a
    function of the number of fermions $N_F$. The long horizontal
    lines correspond to normal clusters, and the short lines to
    polarized clusters. Note that apart from the energy shift, the
    scales of the two figures are the same.}
  \label{energiesp}
   \includegraphics[width=8cm]{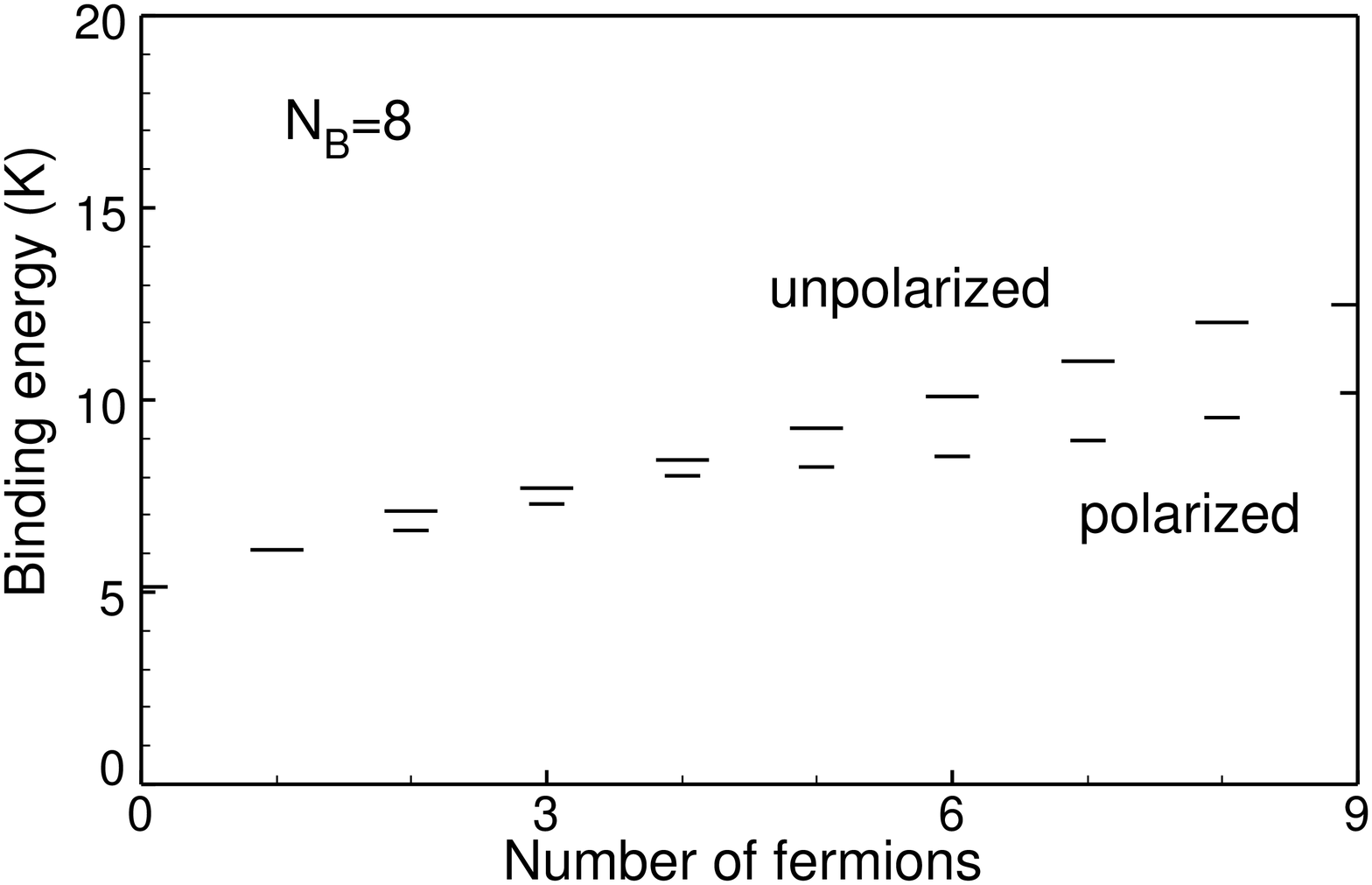}
   \includegraphics[width=8cm]{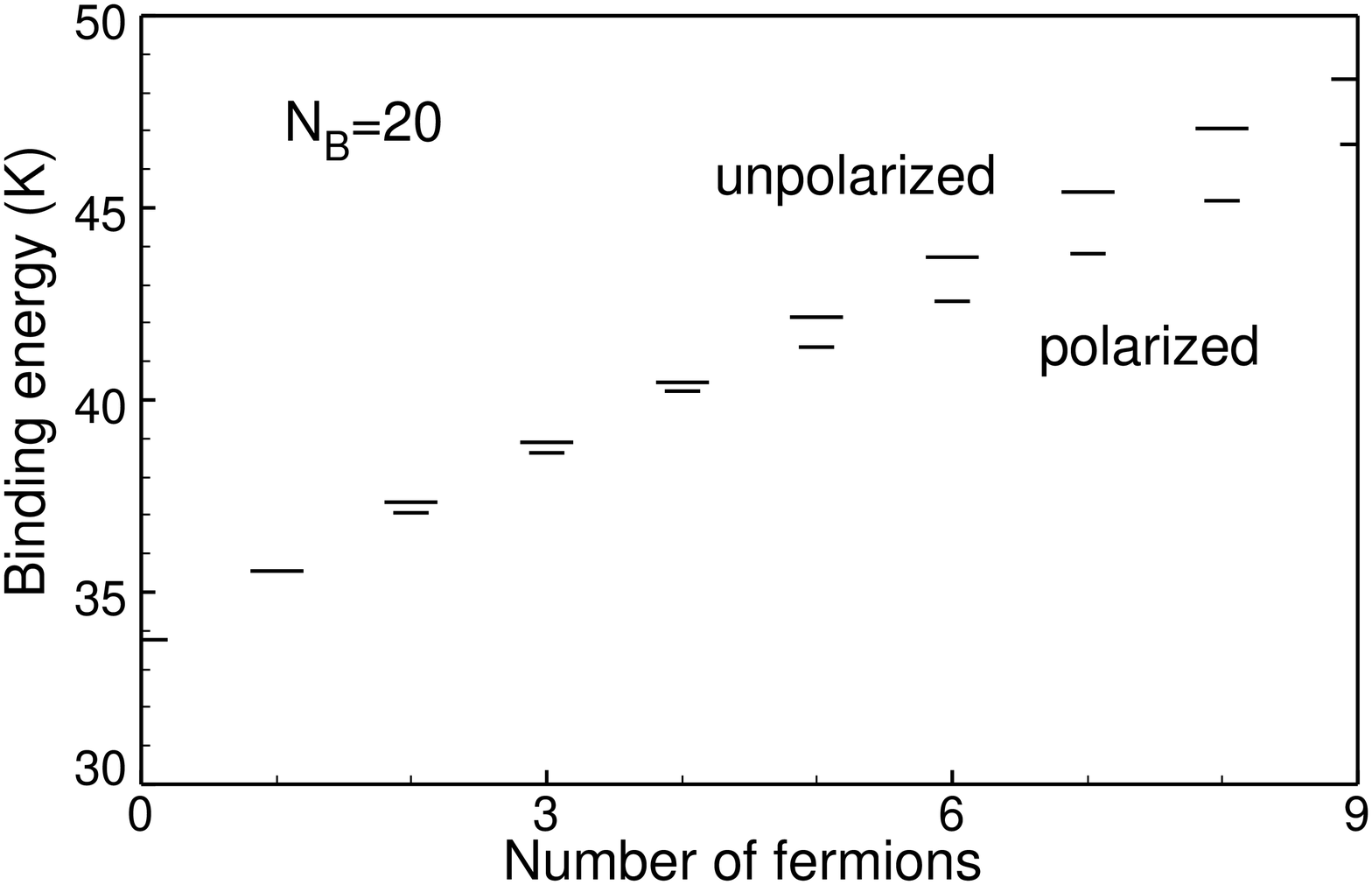}
\end{figure}

\begin{figure}[!htbb]
  \caption{Separation energies, or ionization potentials, (in K) for
    normal $S_{max}$ clusters as a function of the number of fermions
    $N_F$. The lower curve correspond to 8 bosons, and the upper curve
    to 20 bosons.  Dashed lines are just an eye guide.}
  \label{separation}
   \includegraphics[width=8cm]{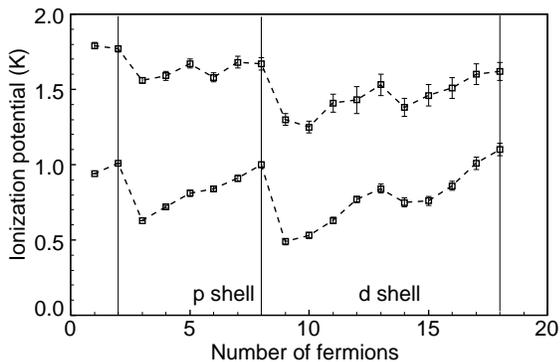}
\end{figure}

The binding energies of Table~\ref{energiesbis}, for normal clusters,
have been plotted in Fig.~\ref{energiesp}, together with those for
polarized clusters [to be discussed in
Section~\ref{sec:polar-mixed-clust} (Table~\ref{enerpol})]. The scale
is such that values for the normal clusters are superimposed.

Basically the energies grow linearly with the number of fermions but
finer details emerge when looking at the fermion chemical potential
\begin{equation}
  \label{chem-pot}
  \mu_F(N_F) = E(N_B, N_F-1) - E(N_B, N_F),
\end{equation}
which is plotted in Fig.~\ref{separation} for two values of $N_B$ and
$N_F\le 18$.  Observe the sudden drop of
$\mu_F$ after $N_F=2$ and 8, corresponding to the closure of
the $1s$ and $1p$ shells.  The relative minimum appearing at $N_F=14$ is
at first somewhat puzzling. Its origin will become clear in
Section~\ref{sec:effect-monop-inter}.

\subsection{The stability map revisited}\label{sec:stab-map-revis-1}

As it has been mentioned in the introduction, one of the most appealing
properties of the mixed He clusters is the existence of {\em instability islands},
namely regions around selected values of $(N_B,N_F)$ in which the system is not bound.
These regions were discovered after Many-Body computations based on self-adjustable
variational functions constructed with Jastrow factors supplemented by 2p-2h and 3p-3h
Configuration Interaction correlations.\cite{guar02} Afterwards, the calculation was refined
by means of the DMC method,\cite{guar03} confirming the previous findings.
Given that both calculations provide really with only upper bounds to the energy,
and having observed the improvement of the present DMC approach, based upon
a different ansatz for the determinantal part of the importance sampling guiding wave function,
we have revised the previous calculation, just to check and eventually
improve the limits of the instability regions.

Indeed, in the previous subsection we have seen that the present level
ordering leads to a noticeable energy gain in the $1d$-shell with
respect to previous works.
 This fact suggests that
some of the clusters previously qualified as {\em metastable}, {\em i.e.}
systems with negative energy but less bound than clusters with a
smaller number of fermions, could be in fact stable. Indeed, a new
computation near the beginning of the $1d$-shell indicates two new
bound systems, namely the clusters $(N_B=3,N_F=11)$ and
$(N_B=4,N_F=9)$. Special attention has been paid to the cluster
$(N_B=1,N_F=18)$, corresponding to full $1s$, $1p$ and $1d$ shells,
but our finding is that this cluster is not bound.

\begin{figure}[h]
  \caption{Stability map of mixed clusters. Solid squares represent
    truly bound states, and open squares represent metastable states.}
  \label{stable}
   \includegraphics[width=8cm]{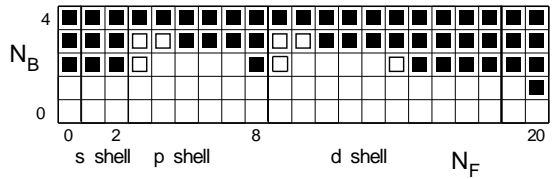}
\end{figure}

The stable clusters are  displayed in Fig.~\ref{stable}, which
supersedes the results previously obtained in
Ref.~\onlinecite{guar03}.
From the experimental point of view, we suggest the study of the regions
$N_B\le 4$ and $N_F\le 5$ to ascertain the stability
limits. The measurements will require an improved mass resolution,
at least 1 amu for clusters up to 25-30 amu.

\subsection{Polarized mixed clusters}\label{sec:polar-mixed-clust}
We have also considered fully polarized mixed systems.  The values
obtained for their binding energies are shown in Table~\ref{enerpol} for
$N_B=8$ and 20. In both cases the number of fermions $N_F \leq 9$ is
limited by the complete filling of the $1s$, $1p$ and $1d$ orbits.
The  asterisk in Table~\ref{enerpol} corresponds to situations in which the
binding energy of the fully polarized system $E_\uparrow(N_B,N_F)$ is
smaller than the energy $E(N_B, N_F-1)$ of the normal cluster with
one fermion less. In the other cases the results correspond to true
bound states.

The energies of polarized clusters are compared with those of
normal clusters in Figure~\ref{energiesp}.
Polarized mixed drops are always less bound than
 the unpolarized cluster with the same number of fermions.
  In other words, they are excited states.

The energy differences between normal an polarized clusters
take  values around 0.25~~K ($N_B=8$) and 0.15~K ($N_B=20$) per fermion.
It is worth mentioning that  theoretical
calculations for  liquid $^3$He provide a difference of  around 0.10~K
per particle at the equilibrium density and $-0.10$~K at densities close
to  the solidification one.~\cite{zong2003}
In other words, the preferred phase
would be the polarized one at high densities.
   This anomalous behavior has been interpreted as
a side effect related to improper nodal surfaces for the unpolarized systems.
The particle density of the fermionic phase in our case is much smaller
than that of the fermionic liquid, and in consequence we cannot ascertain if our
shell-model filling scheme will present such an anomalous crossing
at higher densities.

In the same figure one may
appreciate a sudden change for the differences $\Delta_E =
E_{\rm normal}(N_F) - E_{\rm pol}(N_F)$ at $N_F=4$, which has a
simple interpretation:
it corresponds to the filling of the $1p$ shell for the polarized case
and the related jump in the ionization potential after adding a new fermion.

As for the unbound clusters, all of them have very high spin, and
their binding energy is larger than the energy of the polarized
cluster with one fermion less. Therefore it is very likely that the
set of polarized states corresponds to a stable branch above the
dissociation limit, analogous to the so called {\em displaced terms}
in Atomic Physics.

\begin{table}[!htbb]
  \caption{Binding energies (in K) for fully polarized clusters, for 8 and 20
    bosons. The $^3$He single-particle configuration is indicated in the second
    column. Results marked with an asterisk correspond to clusters with a binding
    smaller than the cluster with a fermion less according to
    Table \ref{energiesbis}.}
  \label{enerpol}
  \begin{tabular}{l|l|ll|l|l}
    \hline
    $N_F$ & Config. & L & S & $N_B=8$ & $N_B=20$ \\
    \hline
    0 &  &  0 & 0 &                 5.14(1)   &33.76(1) \\
    1 & $1s^1$    & 0& 1/2 &        6.08(1)   &35.55(1) \\
    2 & $1s^1 1p^1$& 1 & 1 &        6.65(1)   &37.05(2) \\
    3 & $1s^1 1p^2$& 1 & 3/2&       7.30(1)   &38.61(2) \\
    4 & $1s^1 1p^3$& 0 & 2  &       8.06(1)   &40.22(1) \\
    5 & $1s^1 1p^3 1d^1$& 2 & 5/2 & 8.26(1) * &41.37(2) \\
    6 & $1s^1 1p^3 1d^2$& 3 & 3   & 8.53(1) * &42.57(2) \\
    7 & $1s^1 1p^3 1d^3$& 3 & 7/2 & 8.96(2) * &43.83(3) *\\
    8 & $1s^1 1p^3 1d^4$& 2 & 4   & 9.53(2) * &45.19(3) *\\
    9 & $1s^1 1p^3 1d^5$& 0 & 9/2 &10.19(2) *  &46.63(2) *\\
    \hline
  \end{tabular}
\end{table}

\begin{figure}[!htbb]
  \caption{Separation energies for polarized clusters.}
  \label{separationp}
   \includegraphics[width=8cm]{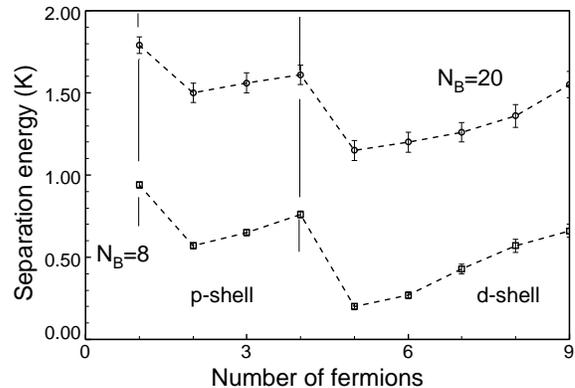}
\end{figure}

The separation energies for the polarized clusters, displayed in
Fig.~\ref{separationp}, follow basically the same pattern than the
separation energies for the normal clusters. Again they recall the
atomic ionization potentials with the sudden drop once a given shell
is closed. Note that for polarized fermions the closure of shells
occurs at $N_F=1$ ($1s$), $N_F=4$ ($1p$) and $N_F=9$ ($1d$).

\section{Effective monopole interaction
  analysis}\label{sec:effect-monop-inter}

In this section we shall treat the results of our calculations as
data, and assume the validity of the shell-model scheme to find an
effective one- plus two-body Hamiltonian that could reproduce them.
This two-body part will only refer to fermions as the boson cluster
 will be assumed as a fixed core that generates the single
particle energies in Table~\ref{previous-calc}, leading to a one body
potential
$$
U = \sum_s n_s \varepsilon_s
$$
where $n_s$ is the number of particles in shell $s$.  The sum is
extended to the occupied shells.

The two-body part is defined by matrix elements
$$
V^{LS}_{rs,tu} = \langle rs:(LS) | V | tu: (LS) \rangle
$$
where $| tu: (LS) \rangle$ is a two-particle state in shells $t$ and
$s$, coupled to orbital angular momentum $L$ and spin $S$, properly
antisymmetrized and normalized.

To simplify matters, the two-body part of the full Hamiltonian will be
separated into monopole ($m$) and multipole ($M$) contributions
$H=H_{mS}+H_M$. The monopole Hamiltonian $H_{mS}$ is defined by the
property of giving the {\em average} energy of configurations at fixed
number of particles $n_r$ and spin $\mathbf {S_r}$. The
closed shells and the one-particle and one-hole states built on them are
configurations with a single state (we call this set $cs\pm 1$).
Hence, their energies are entirely given by $H_{mS}$. For the other
Slater determinants entering our calculations $H_m$ gives an average
value that will be split by the multipole term $H_M$. We shall assume
(and check) that the influence of $H_M$ is small, and simply neglect
it.

The extraction of effective interaction averages goes back in time,
and an important earlier reference is the work of
French.~\cite{French:1969} The form we shall use here came
later~\cite{Abzouzi.Caurier.Zuker:1991} and was used to describe shell
formation properties in nuclear physics.~\cite{Duflo.Zuker:1999}  A
forthcoming review article~\cite{Caurier.Martinez-Pinedo.ea:2003}
contains comprehensive information on formal properties of $H_m$ and
$H_M$.

The average matrix elements are defined as
\begin{eqnarray}\label{eq:centv} \nonumber
  V_{rs}&=&\frac{\sum_{LS} V_{rsrs}^{LS}(2L+1)(2S+1)
    ( 1+(-1)^{L+S}\delta_{rs})}{\sum_{LS} (2L+1)(2S+1)
    ( 1+(-1)^{L+S}\delta_{rs}) }\\
  V_{rs}^S&=&\frac{\sum_L V_{rsrs}^{LS}(2L+1)
    ( 1+(-1)^{L+S}\delta_{rs})}{\sum_L(2L+1)(1+(-1)^{L+S}\delta_{rs})}
  \label{eq:centvs}
\end{eqnarray}
where $V_{rs}$ is the full (scalar) average of two-body matrix
elements, whereas $V^S_{rs}$ are (vector) averages at fixed $S$.  It
is convenient to introduce the following combinations

\begin{eqnarray}
  a_{rs}&=&\frac{1}{ 4}(3V^1_{rs}+V^0_{rs}),\quad
  b_{rs}=V^1_{rs}-V^0_{rs}
  \label{eq:centab}\\
  V_{rs}&=&a_{rs}-\frac{3}{ 4}\,
  \frac{\delta_{rs}}{D_r-1}\,b_{rs}
\end{eqnarray}
where $D_r=2 (2 l_r+1)$ is the maximum number of particles in the
shell. The standard result is then
\begin{eqnarray}
  &&H_{mS} =U+
  \sum_{r\le s} \frac{1}{ (1+\delta_{rs})}\biggl[a_{rs}\,n_r(n_s-\delta_{rs})+
  \nonumber\\
  && +b_{rs}\left({\mathbf S}_r\cdot  {\mathbf S}_s-\frac{3
      n_r}{ 4}\delta_{rs}\right)\biggr],
  \label{eq:hmstan}
\end{eqnarray}
where ${\mathbf S_r}$ is the total spin operator corresponding to the
particles of shell $r$, ${\mathbf S_r}= \sum_{i\in r} {\bf
  \sigma}_i/2$, and the $3n_r/4$ substraction ensures the two-body
nature of $H_{ms}$ by making the spin contribution vanish for {\em
  single particle} states. It has the drawback of producing a non-zero
values at closed shells and single hole states. Therefore it is
preferable to rewrite
\begin{eqnarray}
  &&H_{mS} =U+
  \sum_{r\le s} \frac{1}{ (1+\delta_{rs})}\biggl[V_{rs}\,n_r(n_s-\delta_{rs})+
  \nonumber\\
  && +b_{rs}\left({\mathbf S}_r\cdot  {\mathbf S}_s-\frac{3
      n_r(D_r-n_r)}{ 4(D_r-1)}\delta_{rs}\right)\biggr]
  \label{eq:hms}
\end{eqnarray}
The counter-terms in the second line now ensure its vanishing at the
closed shell as well as at one-particle and one-hole states.  As a
consequence, their energies are fully given by the first line in
Eq.~(\ref{eq:hms}), which we refer to as $H_m$ from now on. The advantage
of this operation is that it decouples the determination of the
$V_{rs}$ and $b_{rs}$ centroids, so that we can proceed with the
former first, as they are the ones that give the global features.

In principle, the six necessary centroids---$V_{ss}$, $V_{sp}$,
$V_{sd}$, $V_{pp}$, $V_{pd}$ and $V_{dd}$---could be extracted from
Table~\ref{two-fermion-8}. However, this parameter-free choice has
large uncertainties and it is better to reserve it as a consistency
check with the results of a more precise fit to the energies in
Table~\ref{energiesbis}, which we call $E_i,\, i=0,18$.  It is
very instructive to start doing the fit by hand, {\em i.e.,} step by step.

Upon filling, the closed shells become new ``cores'': $E_2=E_{Cs},\
E_8=E_{Cp}\ldots$.  The single particle energies are taken from
Table~\ref{previous-calc}. Then

\noindent
$E_2=E_{Cs}=E_0+2\varepsilon_s+V_{ss}$. Extract $V_{ss}$

\noindent
$E_3=E_{Cs}+\epsilon_p+2V_{sp}\equiv E_{Cs}+\bar\epsilon_p$. Extract
$V_{sp}$

\noindent
To extract $V_{pp}$, we do not rely on $E_4$, because it is not purely
given in terms of centroids, but on

\noindent
$E_7=E_{Cs}+5\bar\varepsilon_p+10V_{pp}$ or

\noindent
$E_8=E_{Cp}=E_{Cs}+6\bar\varepsilon_p+15V_{pp}$.

The fit becomes overdetermined, signalling a problem with some basic
assumption about the effective interaction, which we shall try to
identify later. As of now let us settle for a compromise value of
$V_{pp}$. The next step is

\noindent
$E_9=E_{Cp}+\varepsilon_d+2V_{sd}+6V_{pd}\equiv E_{Cp}+\bar
\varepsilon_d$, that determines $V_{sd}+3V_{pd}$. As the two matrix
elements will only appear in this linear combination, the number of
parameters is reduced to five.  Finally, for $V_{dd}$ we have the same
compromise problem we had for $V_{pp}$.  To find reasonable values for
$V_{pp}$ and $V_{dd}$ it was decided to do an overall fit of the five
parameters. In principle, the idea does not seem very sound because
three parameters are {\em apparently} well determined. As we shall
see, this may not be the case, and the numerical fit will turn out to be
sound.

\begin{figure}[t]
  \caption{Comparison of DMC binding energies with the monopole
    Hamiltonian with spin (continuous line) and without spin terms
    (dashed line).  Circles correspond to the computed DMC values for
    normal $S_{max}$ clusters.  The lower line corresponds to the
    fully polarized case, and the upper group to the normal clusters.
    In both cases is $N_B=8$.}
  \label{fig:effective08}
  \includegraphics[width=8cm]{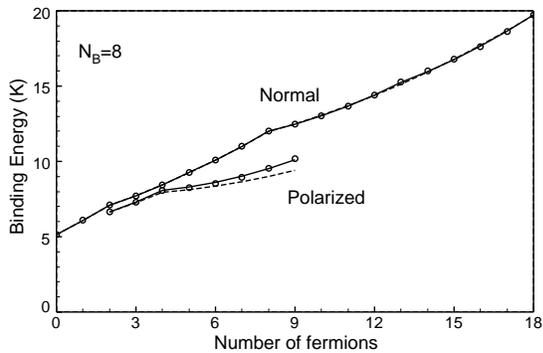}
\end{figure}
\begin{figure}[t]
  \caption{Comparison of calculated separation energies with the
    monopole Hamiltonian with spin ($H_{mS}$, continuous line) and
    without spin terms ($H_{m}$, dashed line), for  $S_{\max}$ states.
     Circles correspond to
    the computed DMC values. The lower group corresponds to $N_B=8$,
    and the upper group to $N_B=20$.}
  \label{fig:sepeff}
  \includegraphics[width=8cm]{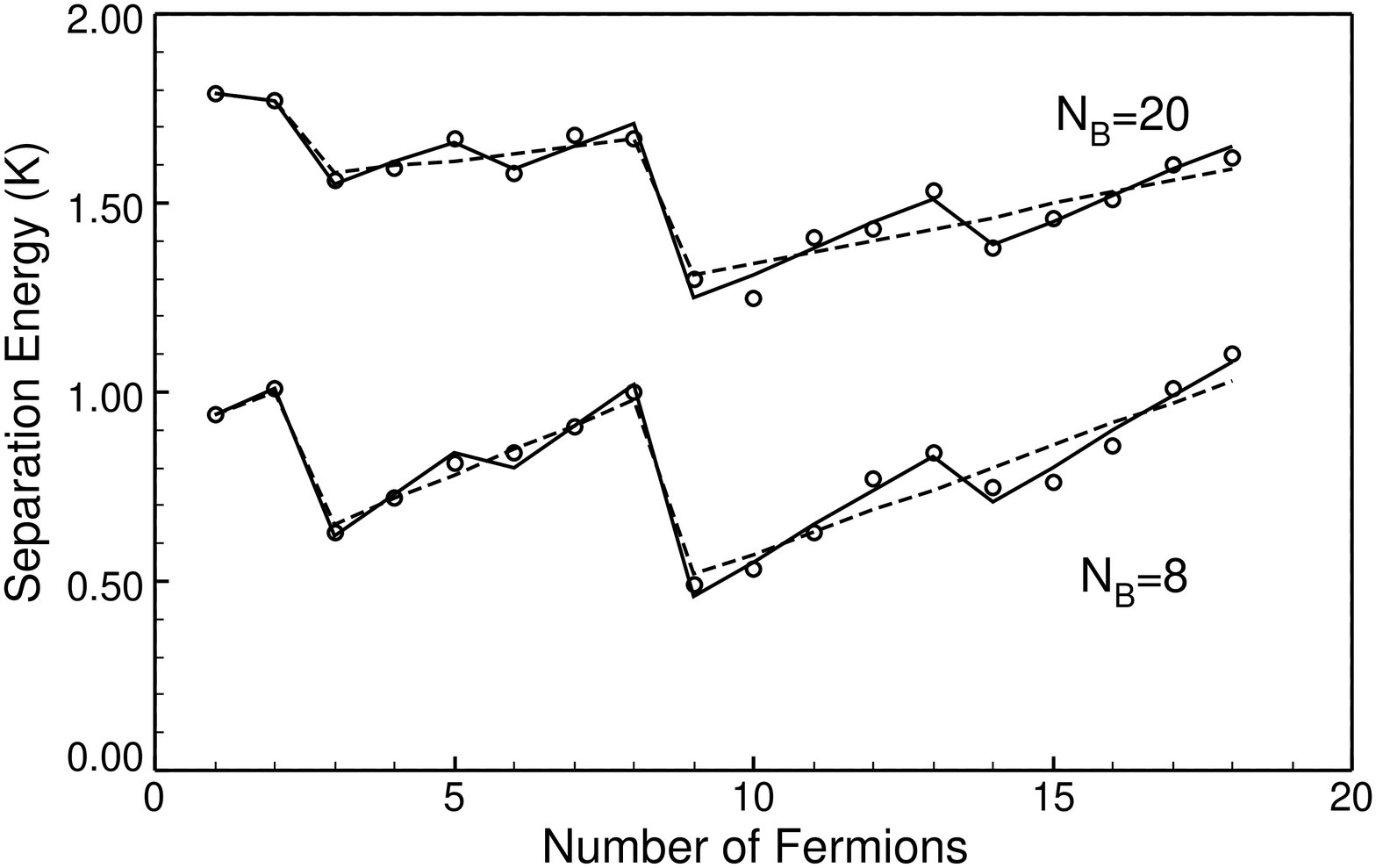}
\end{figure}

The hand-made fit involves $cs\pm 1$ states that are common to the
$S_{max}$ and $L_{max}$ cases in Table~\ref{energiesbis}. In doing the
numerical fit, only the $S_{max}$ states were included. The results
for the binding energies in the $N_B=8$ clusters are given in
Fig.~\ref{fig:effective08}. The agreement is quite excellent, but for
the fully polarized case, also shown, there are significant
discrepancies that can be cured by introducing the full $H_{mS}$
through a single parameter $b_{rs}=b$ so that the contribution of the
second line in Eq.~(\ref{eq:hms}) becomes $b[\, S(S+1)/2-\sum_{r}\, 3
n_r(D_r-n_r)/4(D_r-1)]$. The results of the fit are given in
Table~\ref{table:parameters}. The fitted and calculated curves become
nearly undistinguishable.
\begin{table}[h]
  \caption{The fitted centroids in K}
  \label{table:parameters}
  \begin{tabular}{lrrrr}
    \hline
    & \multicolumn{2}{c}{$N_B=8$} & \multicolumn{2}{c}{$N_B=20$}\\
    \hline
    & Value &    Error  &     Value  &    Error\\
    \hline
    $  V_{ss}$                &  0.073     & 0.006      & -0.019     &   0.007 \\
    $  V_{sp}$                &  0.079     & 0.002      &  0.080     &   0.003 \\
    $  V_{pp}$                &  0.081     & 0.002      &  0.031     &   0.003 \\
    $  (V_{sd}+ 3 V_{pd})/4$  &  0.078     & 0.001      &  0.045     &   0.001 \\
    $  V_{dd}$                &  0.069     & 0.001      &  0.045     &   0.002 \\
    $  b     $                &  0.071     & 0.005      &  0.063     &   0.007 \\
    \hline
  \end{tabular}
\end{table}
As noted at the end of Section~\ref{sec:bind-energ-norm}, the full
energies are rather smooth patterns that tell us little about details.
As a first approximation, Fig.~\ref{fig:effective08} for the normal
$S_{max}$ clusters is reasonably well represented by a straight line,
which would be the analogue of the famous Bethe-Weizs\"acker ``Liquid
drop'' formula for nuclei. The truly sensitive quantities are the
separation energies (chemical potentials) in Eq.~(\ref{chem-pot}) and
Fig.~\ref{separation}. And, indeed, the true test of the monopole
description comes in Fig.~\ref{fig:sepeff}. A smooth linear
approximation to the binding energies would result in a constant. By
introducing $H_m$ there is an enormous improvement in that the shell
effects at closures are well reproduced (dashed line).
However, the more detailed pattern between closures demands the
$S(S+1)$ term in the full $H_{mS}$: the agreement with DMC becomes
truly quantitative (full line).

\begin{figure}[!htbb]
  \caption{Same as Fig.~\ref{fig:sepeff} but for clusters with $L_{\max}$.
}
  \label{fig:sepenerls}
  \includegraphics[width=8cm]{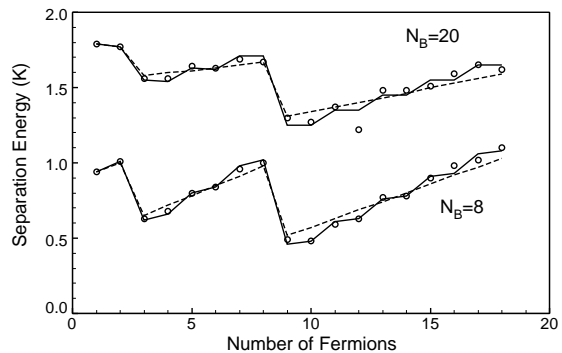}
\end{figure}

The numerical fit was made for the normal case we have called
$S_{max}$, but Table~\ref{energiesbis} contains another normal mode,
$L_{max}$.  As was noted, the hand-made fit is the same for both
couplings, and it gives results that are almost as good as the
numerical fit for $S_{max}$, and very good ones for $L_{max}$. But
here the numerical fit also does a slightly better job, shown in
Fig.~\ref{fig:sepenerls}, where $H_{mS}$ is seen to reproduce
beautifully the staggering pattern between $S=0$ and $S=1/2$ states
(referred to as $L_{max}$ case in Table~\ref{energiesbis}). Here we are
faced with some interesting physics: the numerical fit was chosen to find
good compromise values for $V_{pp}$ and $V_{dd}$, but it does slightly
better than a fit restricted to those over-determined parameters. The
hint is that the other three parameters are not as well determined
as the ``hand-made'' fit suggests. The most likely reason is to
be found in size effects: as fermions are added, the overall radius
evolves, and for a self-bound system it should go asymptotically as
$N_F^{1/3}$.[A strong indication in this sense will be found in
Fig.~\ref{radios8}.] Therefore, the effective matrix elements should
also evolve in a way our simplified $H_{mS}$ ignores: The numerical
fit then emerges as the sound and natural way to define a best
compromise value, not only for $V_{pp}$ and $V_{dd}$ but for all the
monopole parameters.

Finally, Table~\ref{centroids} compares the centroids $V_{rs}^0$ and
$V_{rs}^1$ obtained with the fit to the 19 normal $S_{max}$ clusters
with the values obtained directly from the DMC calculation of
two-fermion states, again with a very good agreement (within the large
errors of about 0.04 K associated to the latter), thus confirming the
consistency of the effective interaction interpretation.
\begin{table}[htb]
  \caption{The two-fermion centroids as obtained from the effective interaction
    (columns labelled Eff) compared with the DMC computed values.}
  \label{centroids}
  \begin{tabular}{|l|rr|rr|}
    \hline
    \multicolumn{5}{|c|}{$N_B=8$}\\
    \hline
    & \multicolumn{2}{|c|}{S=0} & \multicolumn{2}{|c|}{S=1} \\
    \hline
    &  Eff &  DMC  &     Eff &  DMC  \\
    \hline
    ss  &    7.09  &    7.09   &             &           \\
    sp  &    6.57  &    6.61   &      6.64   &   6.65    \\
    pp  &    6.10  &    6.14   &      6.17   &   6.19    \\
    sd  &    5.94  &    5.99   &      6.02   &   6.01    \\
    pd  &    5.46  &    5.49   &      5.54   &   5.51    \\
    dd  &    4.84  &    4.86   &      4.91   &   4.94    \\
    \hline
    \multicolumn{5}{|c|}{$N_B=20$}\\
    \hline
    & \multicolumn{2}{|c|}{S=0} & \multicolumn{2}{|c|}{S=1} \\
    \hline
    &  Eff &  DMC  &    Eff &  DMC  \\
    \hline
    ss  &   37.32   &  37.33    &            &            \\
    sp  &   36.97   &  37.06    &    37.04   &  37.05     \\
    pp  &   36.58   &  36.73    &    36.64   &  36.74     \\
    sd  &   36.44   &  36.47    &    36.50   &  36.49     \\
    pd  &   36.04   &  36.15    &    36.10   &  36.19     \\
    dd  &   35.54   &  35.54    &    35.61   &  35.70     \\
    \hline
  \end{tabular}
\end{table}

It appears that very hard DMC calculations lead to results amenable to
a very simple and cogent interpretation in terms of the monopole
Hamiltonians $H_{mS}$. The $S(S+1)$ contribution is particularly
interesting: Fig.~\ref{fig:effective08} suggests the idea that for
large enough number of fermions the polarized clusters could become
ground states. Though this is only a speculation, it may be also be
taken as a strong invitation to push the study of mixed clusters much further.

\section{The shape of mixed clusters}
\label{sec:shape-mixed-clusters}

\subsection{Normal mixed clusters}\label{sec:norm-mixed-clust}
\begin{figure}[!htb]
  \caption{The values of the root mean square radii (in \AA) for
    bosons (squares) and fermions (circles) referred to the
    center-of-mass of the cluster, as a function of the number of
    fermions $N_F$ in the cluster. The number of bosons are fixed to
    $N_B=8$ and 20.}
  \label{radios8}
  \centerline{
     \includegraphics[width=4cm]{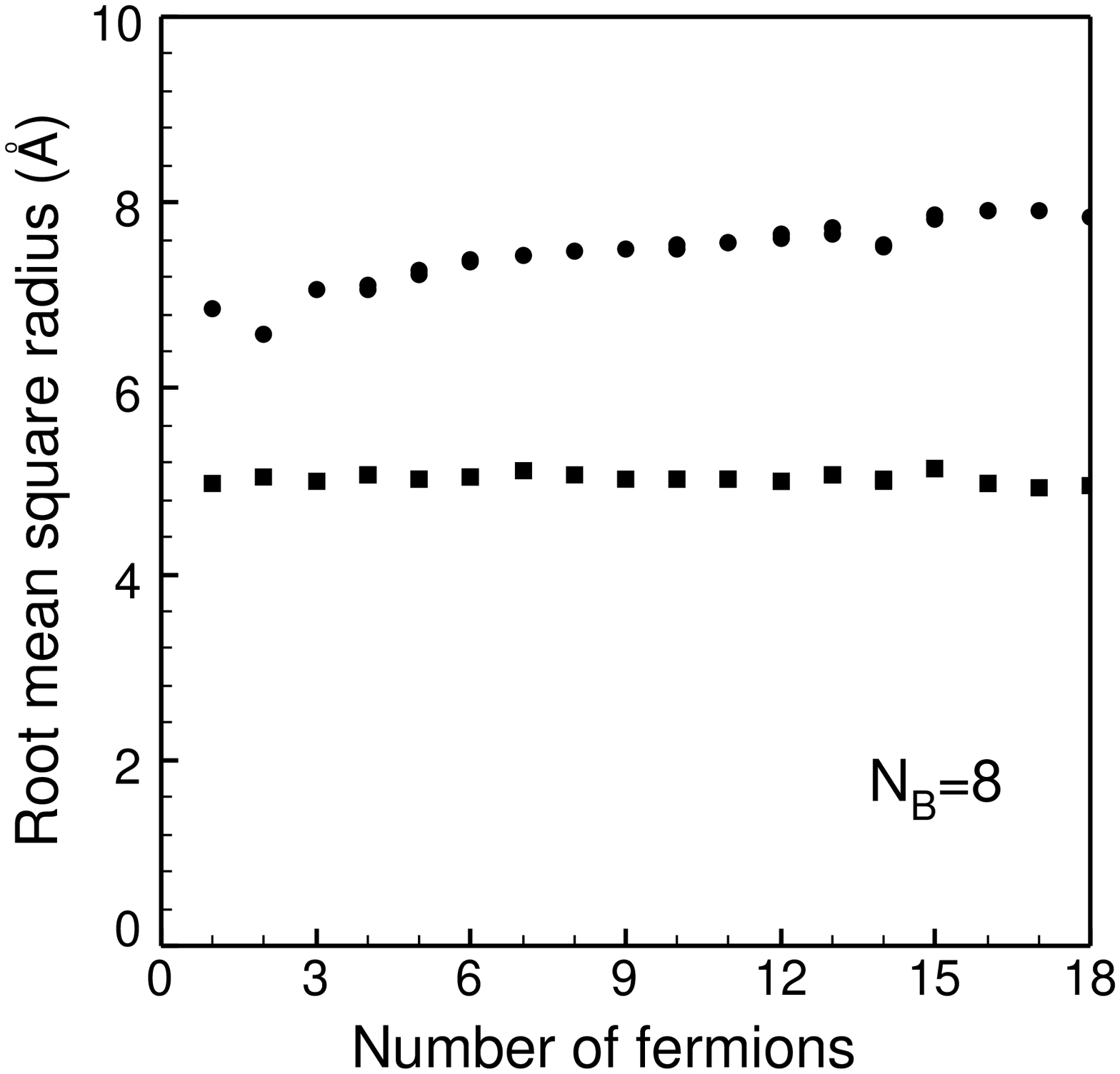}
    \hfill
     \includegraphics[width=4cm]{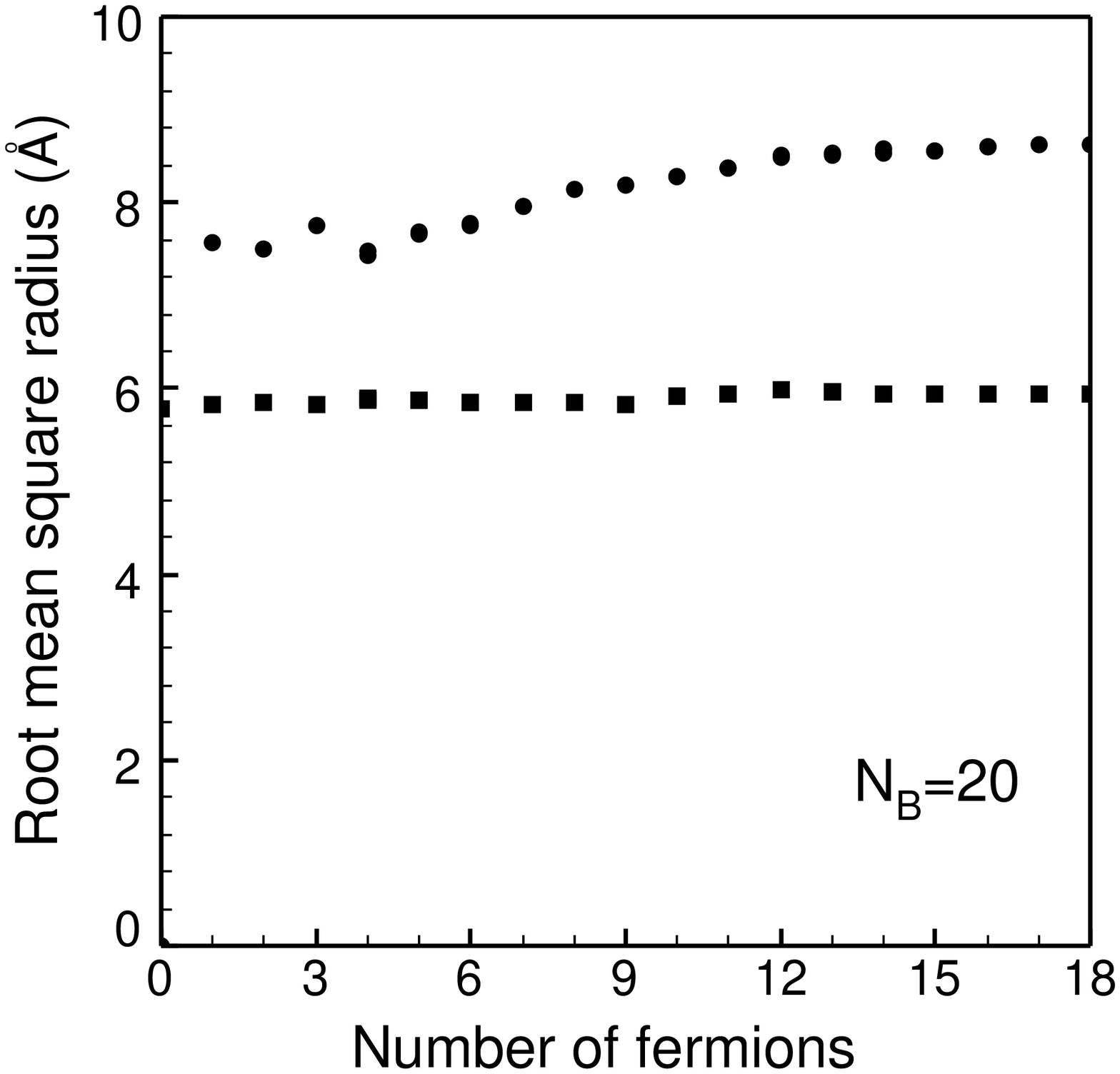}
  }
\end{figure}

In this section we present several figures related to the shape of
mixed clusters. In Fig.~\ref{radios8} there are the values of the root
mean square radii for bosons and fermions referred to the
center-or-mass of the cluster, for the selected cases $N_B=8$ and 20.
There are some fluctuations, probably related to the use of the mixed
estimator method to compute these radii, and thus depending on the
quality of the importance sampling wave function. Apart from these
fluctuations, the most noticeable properties which emerge from these
plots are the almost constancy of the bosonic radii and the smooth
growing of the fermionic radii. This manifests clearly the
representation of the cluster as a quite rigid bosonic core with an
halo of fermions.

\begin{figure}[!htb]
  \caption{The density distributions (in \AA$^{-3}$) of bosons and
    fermions with respect to the center-of-mass of the cluster, for
    the two selected $N_B=8$ and 20 cases.}
  \label{dist8-20}
   \includegraphics[width=6cm]{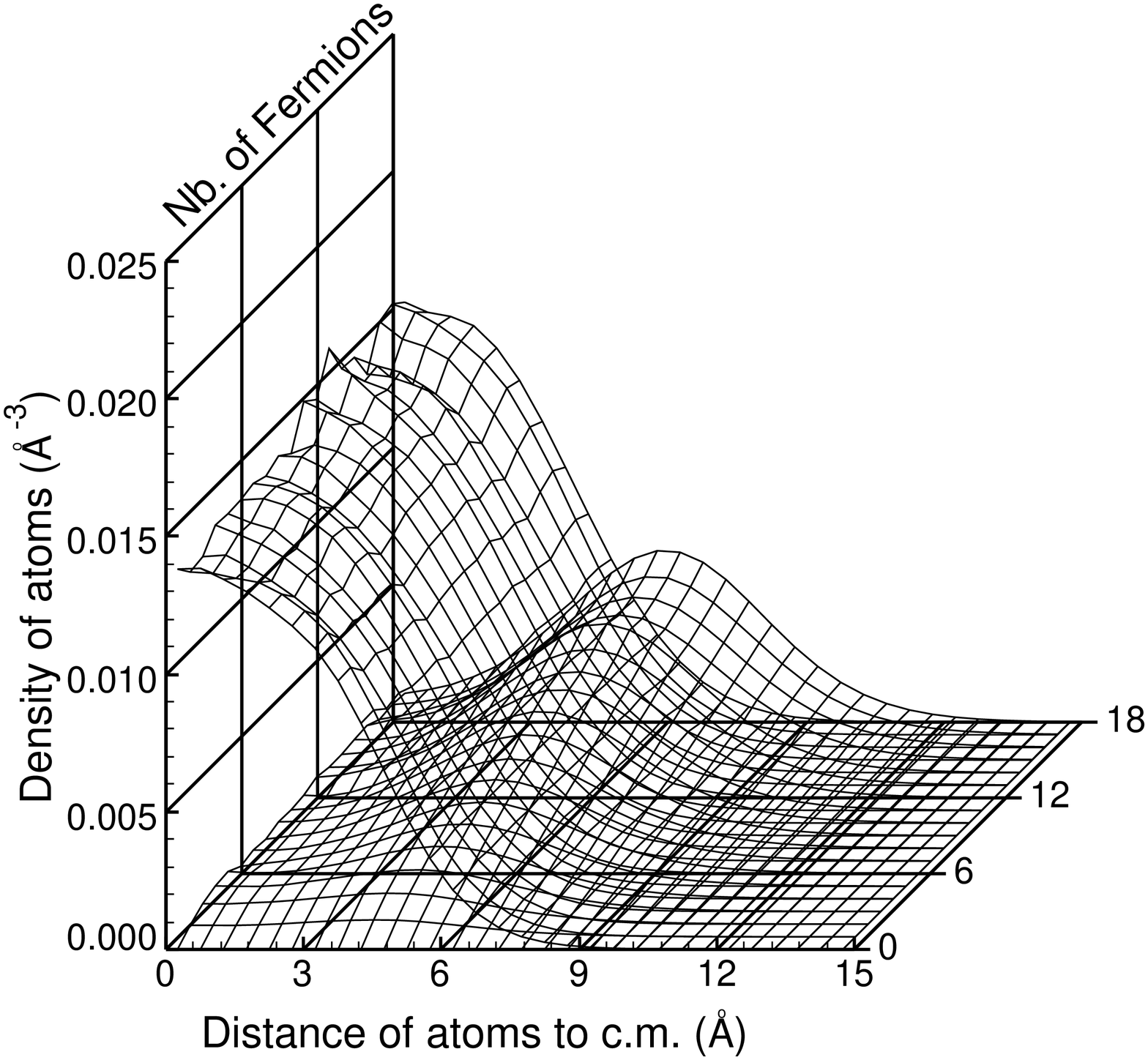}
   \includegraphics[width=6cm]{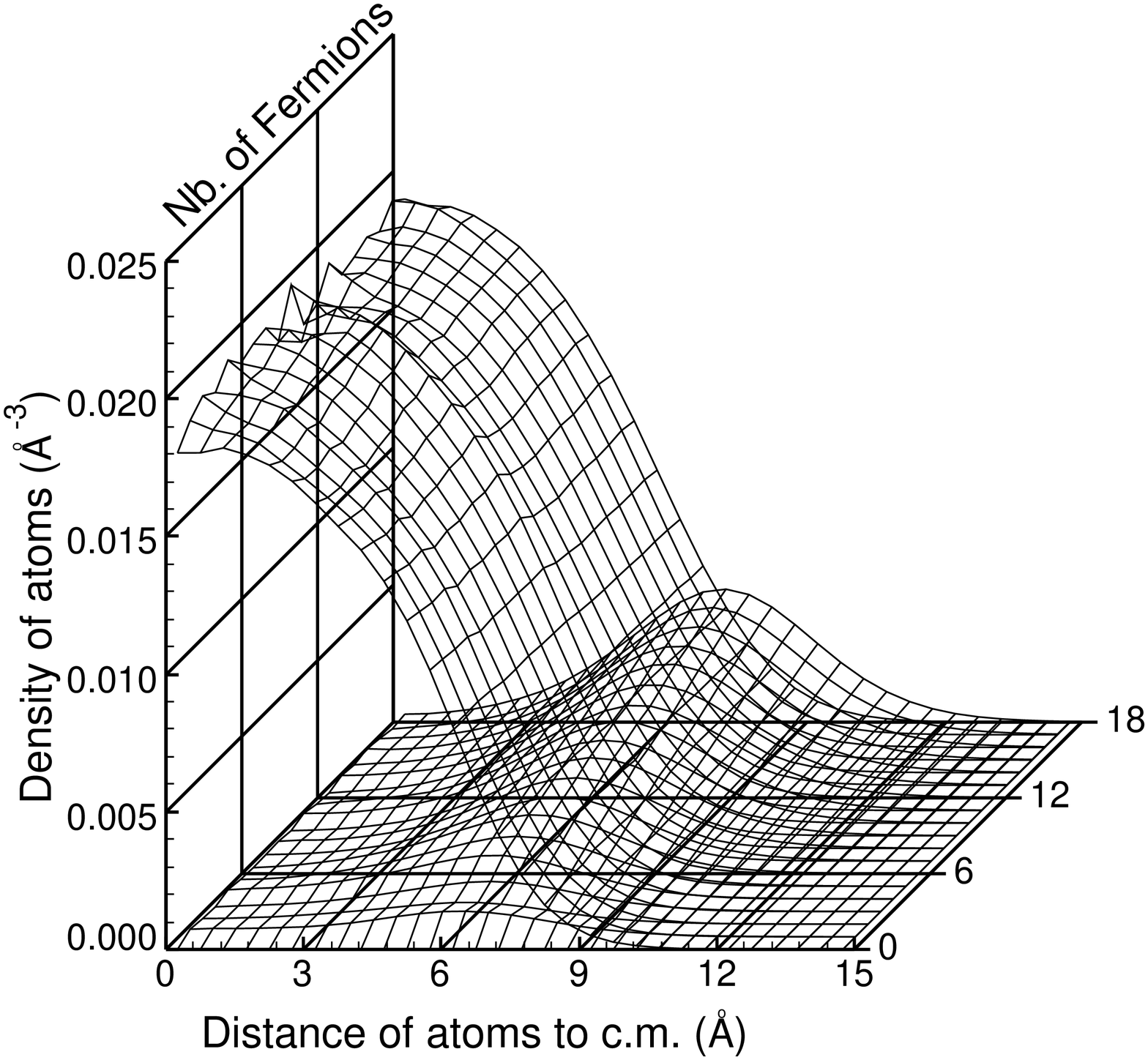}
\end{figure}

This picture is confirmed by the plots of Fig.~\ref{dist8-20}, where
the one-body distributions of bosons and fermions with respect to the
center of mass of the mixed cluster are displayed.  These
distributions are given by
\[\rho_M(r) = \langle \Psi |
\sum_{i=1}^{N_M} \delta ({\bf r} - [{\bf r}_i - {\bf R}] ) |
\Psi\rangle, \] where label $M$ stands for $B$ (bosons) or $F$
(fermions), $N_M$ is the number of atoms of the given species, and $R$
is the center-of-mass of the full drop.  Given that these
distributions may have an angular dependence for open shells, we have
computed the spherical average of them.  The distributions are
normalized to the number of particles of a given species,
\[\int \rho_M(r) d {\bf r} = N_M. \]

We observe in Fig.~\ref{dist8-20} that bosons are located in the same
central region, being slightly compressed as the number of fermions
increases.  This shrinking is more important for the light $N_B=8$
cluster, and is almost negligible when $N_B=20$.  With respect to the
distributions of fermions, they are clearly located at the surface of
the bosonic subcluster, with a small penetration near the center of
the drop in the case of $N_B=8$, more important for larger values of
the number of fermions. In the case of $N_B=20$ the dominating picture
is that of a rigid core of bosons with a fermionic halo.

A complementary information about the shape of clusters is provided by
the two-body distributions, $$ \rho_M({\bf r}, {\bf r'}) =
\frac{2}{N_M (N_M-1)} \langle \Psi | \sum_{i<j}^{N_M} \delta({\bf
  r}-{\bf r}_i) \delta({\bf r}'-{\bf r}_j) | \Psi \rangle, $$ which is
normalized to 1.

Because of the finite size of the system under consideration, this
distribution function depends on two coordinates, ${\bf r}$ and ${\bf
  r}'$, or, equivalently, on the distance of the center-of-mass of the
pair ($({\bf r}+{\bf r}')/2$) to the center-of-mass of the system and
the relative distance (${\bf r}-{\bf r}'$) of the two particles, thus
producing a function very difficult to plot. In order to get a more
friendly quantity we have averaged the above two-body distribution
with respect to its center-of-mass and for the remaining dependence we
have computed the spherical average. The reduced pair distribution so
obtained is now normalized to 1, $$ \int \rho_{12}( r) d {\bf r} = 1.
$$

\begin{figure}[!htb]
  \caption{The boson-boson and fermion-fermion distributions (in
    \AA$^{-3}$), for the two selected cases $N_B=8$ and 20.}
  \label{bbff8-20}
   \includegraphics[width=6cm]{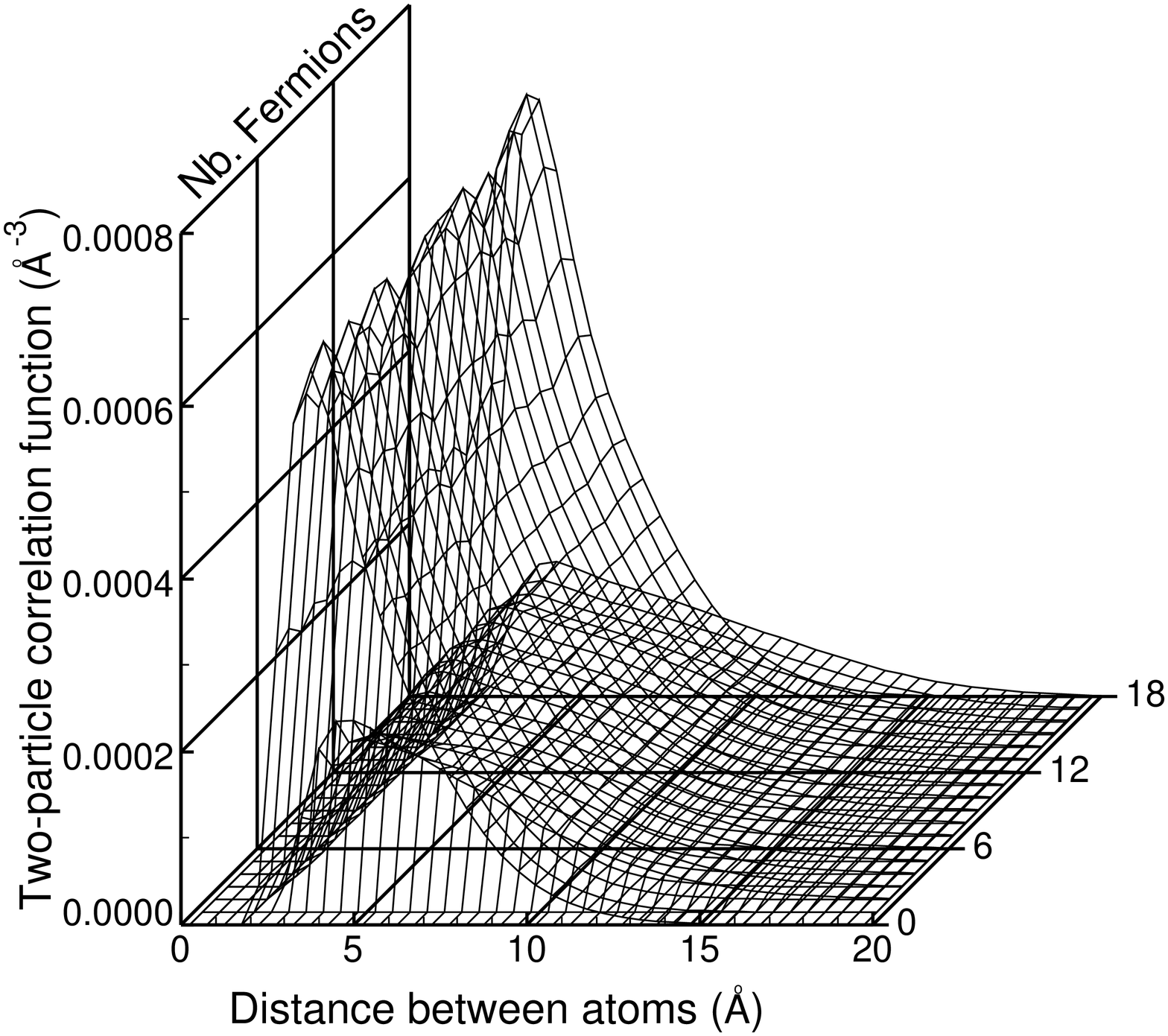}
   \includegraphics[width=6cm]{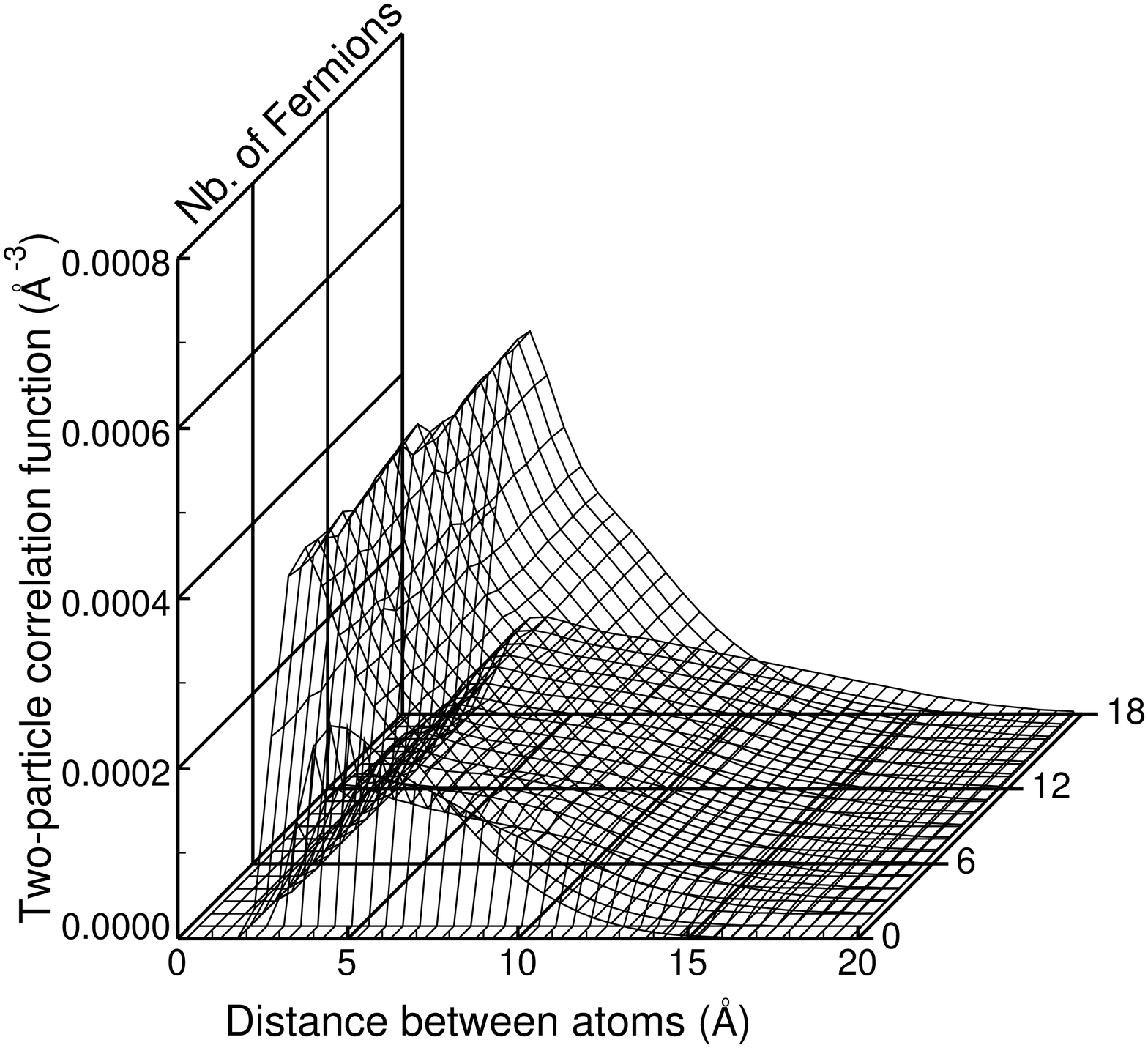}
\end{figure}


\subsection{Polarized mixed clusters}\label{sec:polar-mixed-clust-1}
\begin{figure}[!htb]
  \caption{The values of the root mean square radii (in \AA) referred
    to the center-of-mass, for bosons (filled squares) and fermions
    (filled circles for normal clusters and open circles for polarized
    clusters) as a function of the number of fermions $N_F$ in the
    cluster.  The number of bosons is fixed to $N_B=8$ (left panel)
    and to $20$ (right panel).}
  \label{radpol}
   \centerline{ \includegraphics[width=4cm]{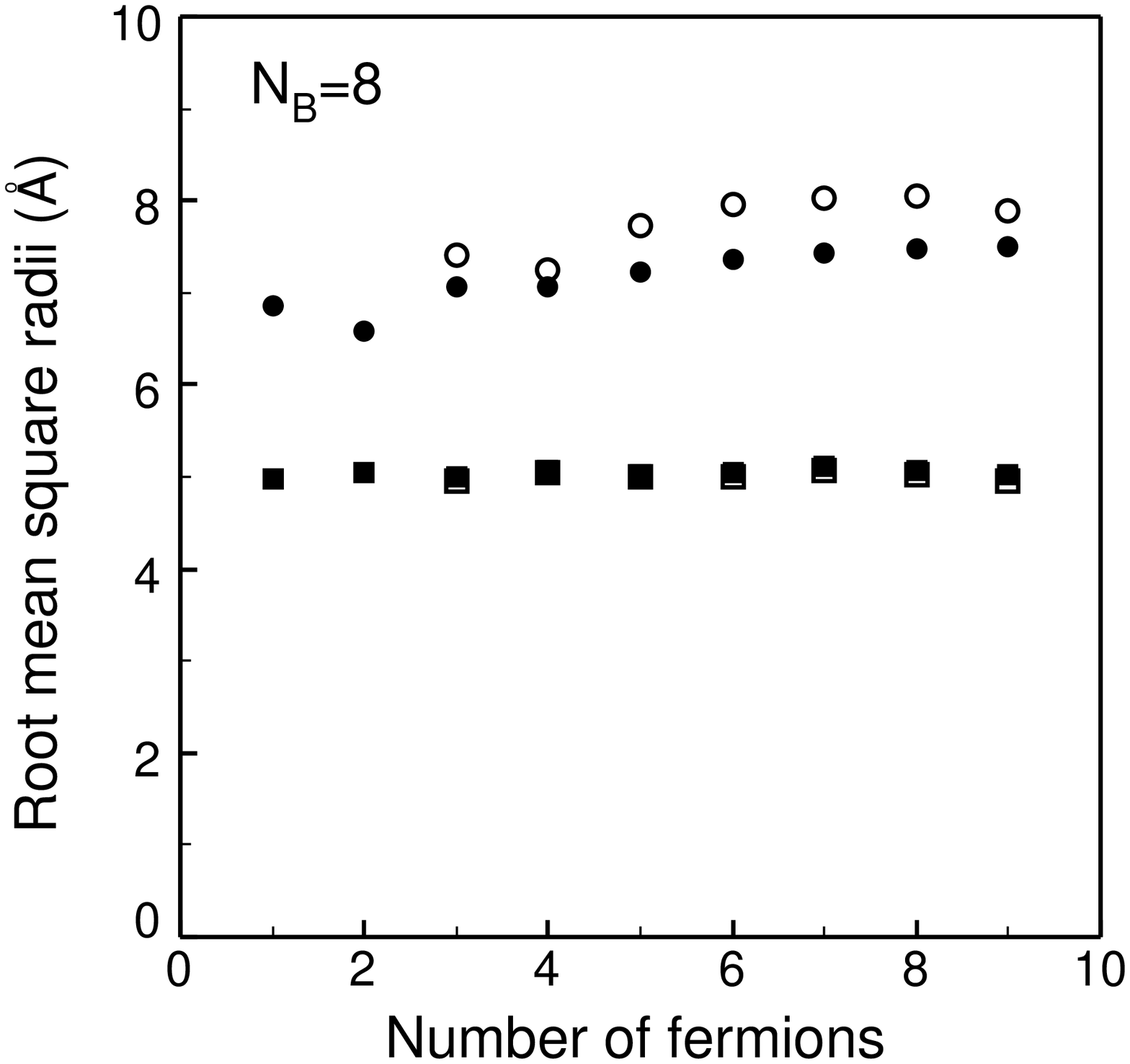}\hspace{1cm}
    \includegraphics[width=4cm]{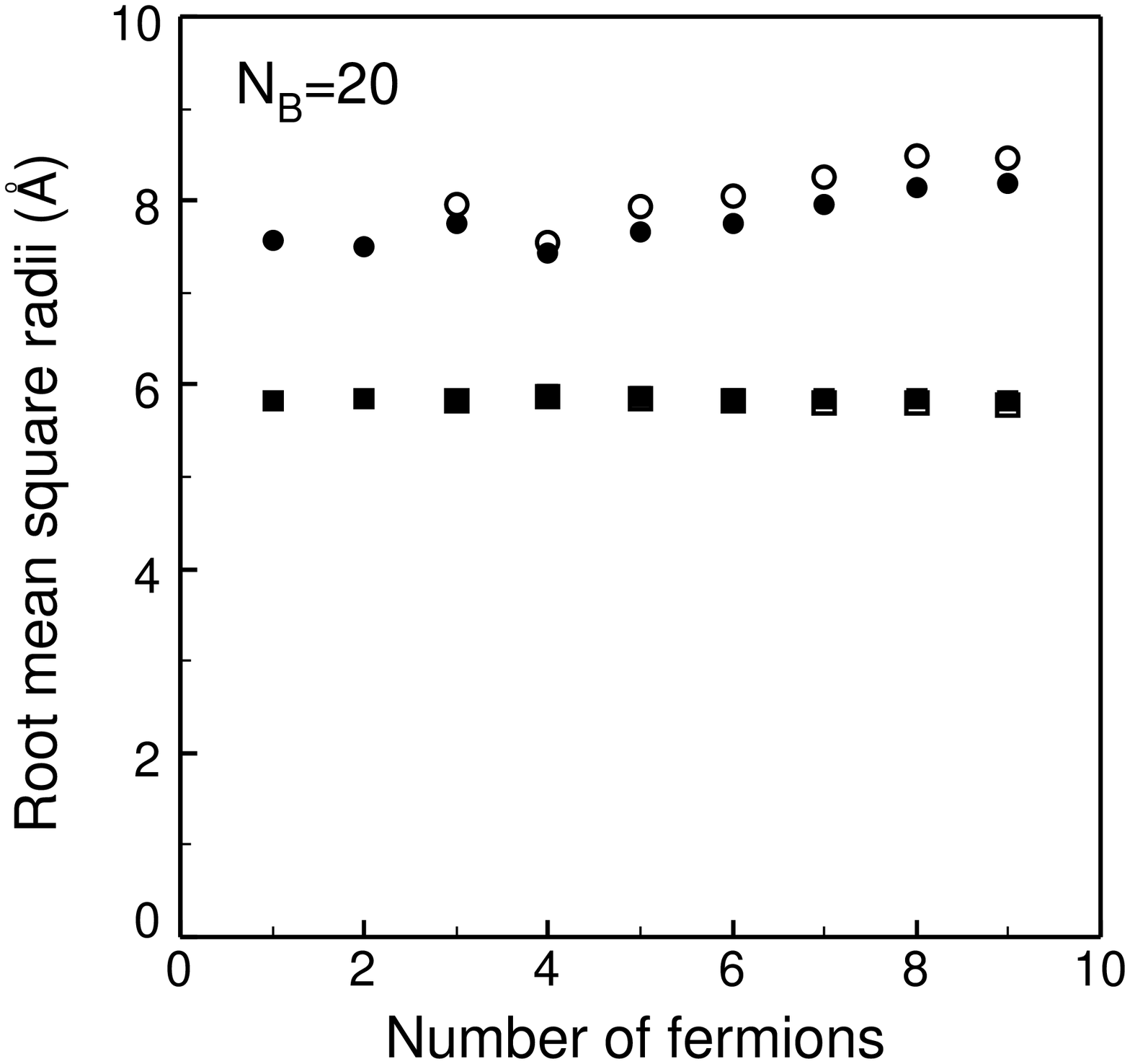}}
\end{figure}

There are noticeable similarities between the distribution functions
related to polarized systems and those corresponding to normal
clusters.  The subcluster of bosons is again {\em hard}, its radius
being practically independent of the number of spin-aligned fermions
$N_F$. Differences with respect to the unpolarized cluster appear when
comparing the root mean square radii of fermions, as shown if
Fig. \ref{radpol} where the fermion halo in the polarized clusters is
larger than in the unpolarized case.

Something similar happens with the density distributions of bosons or
fermions with respect to the center of mass. The former remains
basically unaltered when the number of fermions grows, and the latter
follows the same pattern that in the case of unpolarized clusters.

\begin{figure}[!thb]
  \caption{The fermion-fermion distributions (in \AA$^{-3}$), for
    $N_B=8$ and unpolarized (top) and fully polarized (bottom)
    clusters.}
  \label{bbffp8}
   \includegraphics[width=6cm]{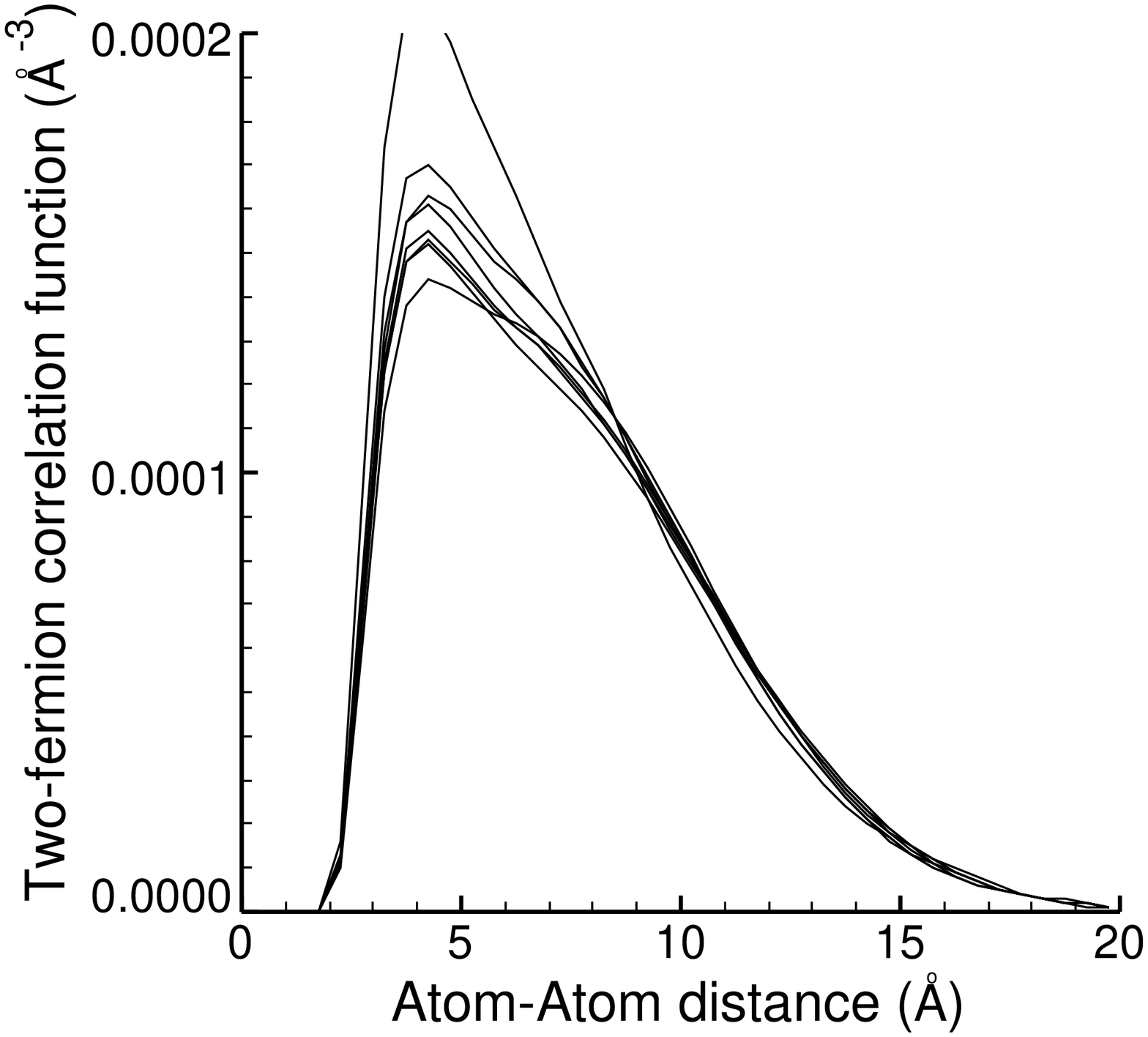}
   \includegraphics[width=6cm]{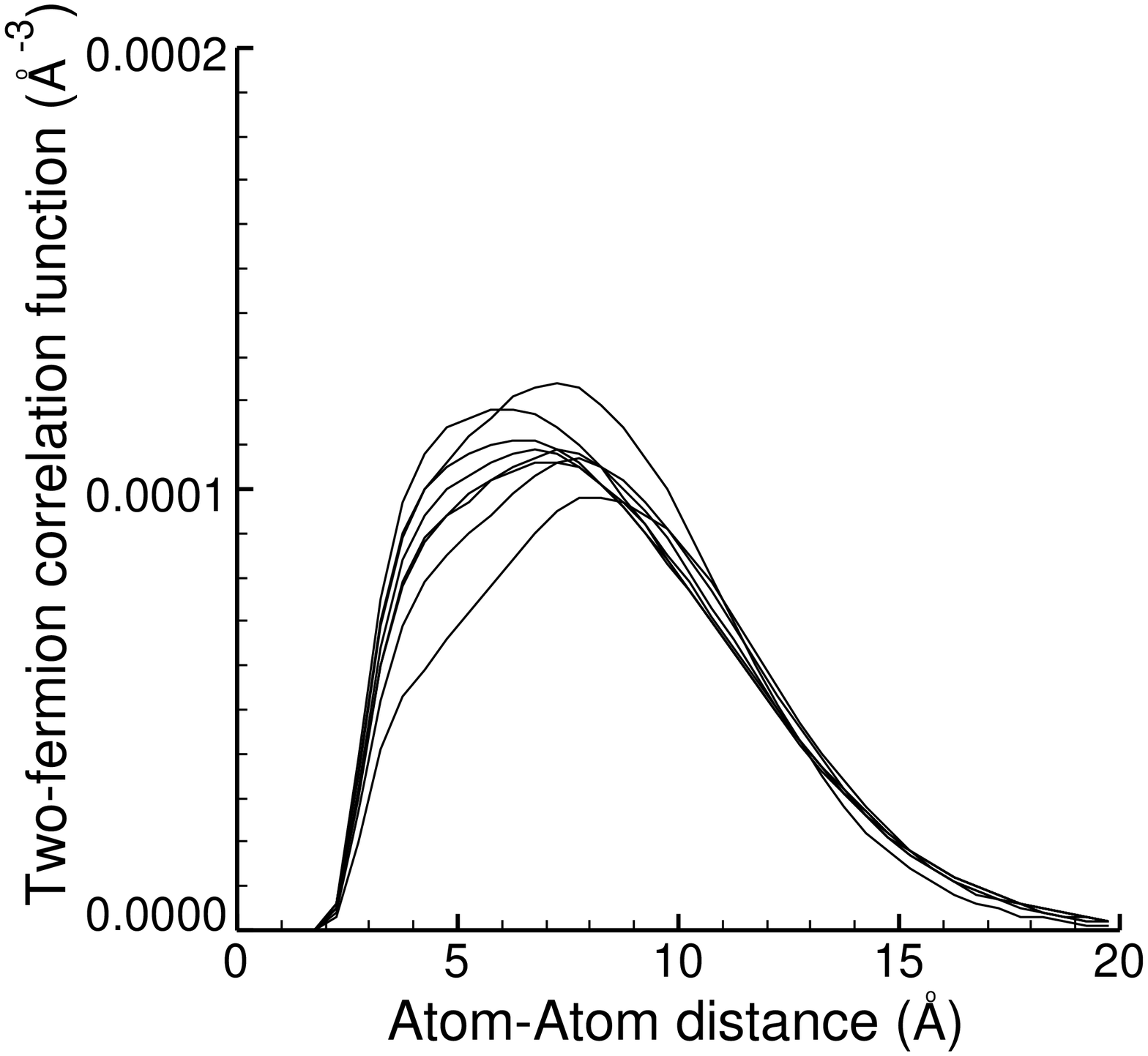}
\end{figure}

There is, however, a remarkable fact in connection with the two-body
distributions and, specifically, the fermion-fermion distributions.
These distributions are shown in Fig.~\ref{bbffp8}, for the two
selected cases $N=8$ and 20. As can be seen there, the fermion-fermion
distributions are very different from those obtained for the normal
systems: the rise near 4~\AA\ is much less pronounced and the range is
much larger, suggesting that the fermions are either less correlated
or subjected to a long-range correlation.

\begin{figure}[!htb]
  \caption{Comparison of the fermion-fermion distributions (lower panel),
    for clusters with $N_B=8$ and $N_F=9$. The dashed line corresponds
    to the polarized cluster and
    continuous line to the normal cluster. }
  \label{polunpol}
   \includegraphics[width=6cm]{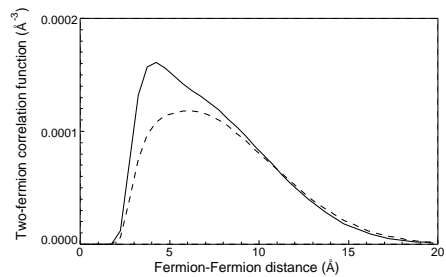}
\end{figure}

In order to appreciate the differences in the density distribution
functions for normal and polarized clusters, we have plotted them in
Figs.~\ref{polunpol} for $N_F=9$ and $N_B=8$. Apart from a clear
change in the shape, the peak of the normal case is close to 4 \AA,
while for the polarized cluster it is close to 7 \AA.

\section{Summary and conclusions}\label{sec:summary-conclusions}
This work has been devoted to a detailed analysis of clusters made of
$^4$He and $^3$He atoms. This new analysis was motivated by the
improved knowledge of the ordering of single-particle orbits, obtained
after a systematic study of the spectrum of a single $^3$He atom bound
to a core of $^4$He atoms,~\cite{fant04} which is expected to provide
the optimal importance sampling trial function for the DMC
calculation.  The study has concentrated on clusters having a
sufficiently large number of $^4$He atoms so as to offer a simplified
pattern: the bosonic constituents arrange themselves as a quite rigid
core whereas the fermionic atoms are distributed in the surface of the
bosonic subcluster, producing a halo. This arrangement, previously
obtained by means of density functional methods, has been confirmed
and pushed down to systems with a small number of constituents.

One of the primary aims was to check the previously obtained stability
map,\cite{guar03} after the optimization of the importance sampling
function, as well as the improvement of the fermionic nodal surfaces.
No significant change occurred, and large instability islands are still
predicted for a small ($N_B\leq 3$) number of $^4$He atoms.

The determination of correlation functions, particularly the
fermion-fermion distribution functions, as well as the analysis in
terms of an effective interaction model suggests that the residual
interactions between the fermions is very weak, of the order of the
computational precision achieved (near 0.1 K). This fact is also
reflected in the insensitivity of the energies to the $(L,S)$ quantum
numbers, the spectrum being essentially determined by the
configuration.

In addition to the {\em normal} fermionic phase, with a small value of
the spin, we have also analyzed the possibility of having a {\em
  ferromagnetic} phase, with all spins aligned. The normal phase is
energetically favored but the ferromagnetic one gives rise to bound
states, even for a moderately large number of $^3$He atoms. For
example, a cluster with 20 bosons is able to bind up to six fermions,
but beyond that number the system is above the dissociation limit.
Nevertheless, because of the large values of the spin for the
ferromagnetic phase, one may expect these unbound states to be
long-lived, like in the case of
polarized liquid $^3$He,  and thus to be experimentally detectable.
Perhaps by circulating
bosonic clusters through a cold atmosphere of polarized $^3$He atoms
one could create these spin-aligned states sticking $^3$He atoms
one-by-one to the bosonic seed.

The  mixed systems may
have a very rich excitation spectrum, because of the gap between the
normal and the polarized phase. For example, for $N_B=20$ and $N_F=6$
there is a difference of 1 K between the normal state (configuration
$1s^2 1p^4$) and the polarized state (configuration $1s^1 1p^3 1d^2$).
The analysis of the intermediate filled configurations, such as $1s^2
1p^3 1d^1$ or $1s^2 1p^2 1d^2$ puts heavy demands on the DMC algorithm
but it would be relatively simple in the density-functional method.
Though it does not seem possible with the present experimental
techniques to measure the spectrum, it is worth remembering that the
excitation spectrum plays a relevant role in the production abundances
of clusters.\cite{bruch02, guar04}

Finally one should stress the power of the effective monopole
interaction analysis that points to the basic simplicity of the DMC
results and invites to take on the challenge of unearthing the deep
reasons of such simplicity.

\acknowledgments This work has been supported by MCyT/FEDER (Spain),
grant number FIS2004-00912, GV (Spain), grant number GV2003-002 and
MIUR (Italy), cofin-2001025498. Part of this study was done
while RG was visiting professor at Université Louis Pasteur in Strasbourg.
One of us (RG) acknowledges financial
support of the Secretar\'{\i}a de Estado de Educaci\'on y
Universidades (Spain), Ref. PR2003-0374, as well as DEMOCRITOS by his
hospitality.

\end{document}